\documentclass[11pt,a4paper]{article}

\usepackage{jheppub}
\usepackage{t1enc}
\usepackage{mathrsfs}
\usepackage[utf8]{inputenc}
\usepackage{afterpage}
\usepackage{amsmath}
\usepackage{amssymb}
\usepackage{bbm}
\usepackage{multirow}
\usepackage{booktabs}
\usepackage{subfig}
\usepackage{hyperref}
\usepackage{floatrow}

\newcommand{\MSb}{\overline{\textrm{MS}}}

\title{Quark mass anomalous dimension and $\Lambda_{\MSb}$\\from the twisted mass Dirac operator
spectrum}

\author[a,b]{Krzysztof Cichy}

\affiliation[a]{NIC, DESY, Platanenallee 6, 15738 Zeuthen, Germany}
\affiliation[b]{Adam Mickiewicz University, Faculty of Physics,
Umultowska 85, 61-614 Poznan, Poland}

\emailAdd{krzysztof.cichy@desy.de}

\preprint{DESY 13-212, SFB/CPP-13-91}

\abstract{
We investigate whether it is possible to extract the quark mass anomalous dimension and its scale
dependence from the spectrum of the twisted mass Dirac operator in Lattice QCD.
The answer to this question appears to be positive, provided that one goes to large enough
eigenvalues, sufficiently above the non-perturbative regime.
The obtained results are compared to continuum perturbation theory.
By analyzing possible sources of systematic effects, we find the domain of applicability of the
approach, extending from an energy scale of around 1.5 to 4 GeV.
The lower limit is dictated by physics (non-perturbative effects at low energies), while the upper
bound is set by the ultraviolet cut-off of present-day lattice simulations.
The information about the scale dependence of the anomalous dimension allows also to extract the value
of the $\Lambda_{\MSb}$-parameter of \mbox{2-flavour} QCD, yielding the
value $303(13)(25)$ MeV, where the first error is statistical and the second one
systematic.
We use gauge field configuration ensembles generated by the European Twisted Mass Collaboration
(ETMC) with 2 flavours of dynamical twisted mass quarks, at 4 lattice spacings in the range between
around 0.04 and 0.08 fm.
}

\begin{document}
\maketitle

\section{Introduction}
\label{sec:intro}
The spectrum of the Dirac operator in Lattice QCD provides very interesting information
about several important properties of QCD.
In particular, the low modes of the Dirac operator are intimately linked to the chiral condensate,
the order parameter of spontaneous chiral symmetry breaking, via the Banks-Casher relation
\cite{Banks:1979yr}. This relation has recently led Giusti and L\"uscher \cite{Giusti:2008vb} to a
new method of extracting the chiral condensate -- from the slope of the mode number
$\nu(M)$, which counts the number of eigenmodes of the Hermitian Dirac operator $D^\dagger D$ below
some threshold value $M^2$. The mode number can be very efficiently evaluated using a stochastic
method, the so-called spectral projector method. Using this method, we have recently calculated the
chiral condensate in the continuum limit, with 2 and 2+1+1 flavours of dynamical twisted mass quarks
\cite{Cichy:2013gja,Cichy:2013egr}.
A similar study using $N_f=2$ clover fermions was reported in Ref.~\cite{Engel:2013rwa}.
Another way of linking the low Dirac eigenmodes to the chiral condensate is provided by chiral random
matrix theory
\cite{Shuryak:1992pi,Damgaard:2000qt,DeGrand:2005vb,Lang:2006ab,Fukaya:2007fb,Fukaya:2007yv,
Fukaya:2010na, Bernardoni:2010nf,Splittorff:2012gp}.
Apart from the link to the chiral condensate, the Dirac spectrum is also related to the
topological susceptibility \cite{Giusti:2008vb,Luscher:2010ik,Cichy:2011an,Cichy:2013lat,Cichy:2013rra}
and the quark mass
anomalous dimension. This has become an important tool for extracting the latter at the infrared (IR)
fixed point in conformal field theories
\cite{DeGrand:2009hu,DelDebbio:2010ze,Cheng:2011ic,Hasenfratz:2012fp,Cheng:2013eu,
Cheng:2013bca,Patella:2011jr,Patella:2012da,Keegan:2012xq,Perez:2013tnv,Landa-Marban:2013oia}.

The issue of the scale dependence (or running coupling dependence) of quantities like the anomalous
dimensions and renormalization constants is most often addressed on the lattice using the Schr\"odinger
functional, combined with
step scaling techniques \cite{Luscher:1991wu,Luscher:1992an} -- for applications to the
computation of the running quark mass (governed by the quark mass anomalous dimension) see e.g.
Refs.~\cite{Capitani:1998mq,DellaMorte:2005kg,Bursa:2009we,Fritzsch:2010aw,Aoki:2010wm,DeGrand:2010na,
Svetitsky:2010zd,DeGrand:2012qa,Lopez:2012mc}.
However, step scaling methods can also be used to extract the scale dependence of the
renormalized quark mass in the RI-MOM scheme -- see
e.g. Refs.~\cite{Zhestkov:2001hx,Constantinou:2010gr,Arthur:2010ht,Arthur:2012opa}.

In this paper, we want to investigate whether it is possible to extract the implicit scale dependence of
the
quark mass anomalous dimension from the spectrum of the Dirac operator in 2-flavour Lattice
QCD.
In general, any comparison of lattice-extracted quantities with perturbation theory (PT) needs the
existence of an energy scale window such that this scale $\mu$ is:
\begin{enumerate}
 \item high enough for PT to be applicable, i.e. $\mu$ should be much larger than a
typical low
energy QCD scale of the order of a few hundred MeV ($\mu\gg\mathcal{O}(\Lambda_{\rm QCD})$),
 \item low enough to avoid large cut-off effects, i.e. $\mu\ll\Lambda_{\rm lat}$, where $\Lambda_{\rm
lat}$ is the lattice ultraviolet cut-off (inverse lattice spacing).
\end{enumerate}
The present-day lattice simulations are typically performed with lattice spacings between around 0.05
and 0.15 fm, which corresponds to cut-offs of ca. 1.3 to 4 GeV.
This means that the lattice window for establishing contact with perturbation is very narrow and even
its very existence is limited to the finer lattice spacings reached nowadays.
Our aim is to investigate whether such a window exists for the quark mass anomalous dimension and
whether the latter can be accessed with methods of Refs.
\cite{DeGrand:2009hu,DelDebbio:2010ze,Cheng:2011ic,Hasenfratz:2012fp,Cheng:2013eu,
Cheng:2013bca,Patella:2011jr,Patella:2012da,Keegan:2012xq,Perez:2013tnv,Landa-Marban:2013oia}.
If the answer is positive, the method can allow for a relatively cheap extraction of the running of the
quark mass anomalous dimension.
As such, it can at least complement more standard approaches to the computation of this quantity on the
lattice, e.g. in the framework of the Schr\"odinger
functional~\cite{Capitani:1998mq,DellaMorte:2005kg,Bursa:2009we,Fritzsch:2010aw,Aoki:2010wm,
DeGrand:2010na,Svetitsky:2010zd,DeGrand:2012qa,Lopez:2012mc} or the RI-MOM method
\cite{Zhestkov:2001hx,Constantinou:2010gr,Arthur:2010ht,Arthur:2012opa}.
In addition, the information about the scale dependence of the renormalized quark mass allows to extract
the $\Lambda_{\MSb}$-parameter of the theory, by matching of the continuum extrapolated lattice data to
4-loop PT.

The paper is organized as follows. In section \ref{sec:theory}, we describe the theoretical principles
of the employed method and discuss its potential limitations. Section \ref{sec:fit} presents our analysis
strategies and section \ref{sec:setup} the lattice setup. In
section \ref{sec:results}, we show our results. Section \ref{Sec:conclusions} concludes. Additional
tests are presented in 4 appendices.

\section{Theoretical principles}
\label{sec:theory}

\subsection{Spectral density and mode number of the Dirac operator}
\label{sec:spectral}
The main aim of this paper is to check if it is possible to extract the running of the quark mass
anomalous dimension from the lattice QCD Dirac operator spectrum. The formula that will be used in the
numerical part can not be derived from first principles. Therefore, we will present here the
arguments that lead to it, coming from different premises: perturbation theory and renormalization group.

\subsubsection*{Scaling of the spectral density in theories with an infrared fixed point}
The scaling of the spectral density of the Dirac operator $\rho(\lambda)$ (where its
eigenvalues are denoted by $\lambda$) is related to the quark mass anomalous dimension.
Let us start with a short r\'esum\'e of the situation in gauge theories with an IR fixed
point. Using the properties of these systems, it is possible to show that the scaling of the spectral
density of the Dirac operator is related to the scheme-independent mass anomalous dimension at the
fixed point \cite{DeGrand:2009hu,DelDebbio:2010ze,Patella:2011jr}.
This relation can be written in the following form: \cite{Patella:2012da}
\begin{equation}
\label{eq:spectral}
 \rho(\lambda) = \hat\rho_0 \, \mu^{\frac{4\gamma_m^*}{1+\gamma_m^*}} \,
\lambda^{\frac{3-\gamma_m^*}{1+\gamma_m^*}},
\end{equation}
at leading order, where: $\hat\rho_0$  -- dimensionless constant, $\mu$  -- renormalization
scale, $\gamma_m^*$ -- quark mass anomalous dimension at the
IR fixed point.
This can be rewritten using the integrated spectral density, i.e. the (dimensionless) mode number:
\begin{equation}
\label{eq:nu1}
\nu_R(M_R)=2V\int_0^{M_R}
d\lambda\,\rho(\lambda)=\hat\rho\mu^{\frac{4\gamma_m^*}{1+\gamma_m^*}} \,
M_R^{\frac{4}{1+\gamma_m^*}},
\end{equation}
where $\nu_R(M_R)$ is the renormalized number of eigenmodes of the Hermitian Dirac operator
$D^\dagger D$ below some renormalized threshold eigenvalue $M_R^2$, $V$ is the volume
and $\hat\rho$ is a constant with the dimension of volume.
It was shown in Ref.~\cite{Giusti:2008vb} that the mode number is renormalization group invariant,
i.e. $\nu_R(M_R)=\nu(M)$.

Since the mode number is a quantity easily accessible on the lattice, it is, in principle,
possible to use the above formula to extract the quark mass anomalous dimension at the IR
fixed point.
However, in actual lattice simulations, scale invariance is broken by a non-zero quark mass $m$
\footnote{This is in sharp contrast to QCD-like theories, where chiral symmetry is broken spontaneously
(even at zero quark mass), as well as explicitly (by the non-zero quark mass).},
leading to the development of a fermion condensate and a mass gap \cite{DelDebbio:2010ze}.
The lattice results can then be described within mass-deformed conformal gauge theory.
In the presence of a non-zero quark mass, the mode number equation \eqref{eq:nu1} is modified to:
\cite{Patella:2012da}
\begin{eqnarray}
\label{eq:nu2}
\nu_R(M_R)&=&2V\int_0^{\sqrt{M_R^2-m_R^2}}d\lambda\,\rho(\lambda) =
2V\int_0^{\lambda_{IR}} d\lambda\,\rho(\lambda)
+ 2V\int_{\lambda_{IR}}^{\sqrt{M_R^2-m_R^2}} d\lambda\,\rho(\lambda) = \nonumber\\
&\approx&\nu_0(m_R)+\hat\rho\mu^{\frac{4\gamma_m^*}{1+\gamma_m^*}} \,
M_R^{\frac{4}{1+\gamma_m^*}},
\end{eqnarray}
where $m_R$ denotes the renormalized quark mass and $\lambda_{IR}$ is some infrared scale below which
quark mass effects are relevant.
The mass deformation introduces an extra term $\nu_0(m_R)$, dependent solely on the quark mass.
Its practical consequence is that the anomalous dimension at the fixed point $\gamma_m^*$ can not be
extracted for too small values of the scale $M_R$, where effects of the mass deformation can be
large. However, one can get control over these effects by including the term $\nu_0(m_R)$
in fits and simulating at more than one light quark mass, to explicitly check its influence.

\subsubsection*{Scaling of the spectral density in chirally broken theories}
The situation is
somewhat similar in chirally broken systems, like QCD. The infrared behaviour
of such systems is very
different from theories with an IR fixed point. However, the effects of spontaneous chiral
symmetry breaking should be relevant only below some scale $\Lambda_\chi$.
Hence, one can conjecture
that well above this scale, the scaling of the mode number can resemble the one in conformal systems,
making Eq.~\eqref{eq:nu2} valid.
However, since no infrared fixed point occurs in chirally broken systems, the quark mass anomalous
dimension at the IR fixed point $\gamma_m^*$ is replaced by an anomalous
dimension $\gamma_m(M_R)$ dependent on the running coupling (and hence implicitly scale-dependent)  and
Eq.~\eqref{eq:nu2} becomes:
\begin{equation}
\label{eq:nu3}
\nu_R(M_R)=\nu_0(m_R)+\hat\rho\mu^{\frac{4\gamma_m(M_R)}{1+\gamma_m(M_R)}} \,
M_R^{\frac{4}{1+\gamma_m(M_R)}},
\end{equation} 
for $M_R\gg\Lambda_\chi$.
Such an approach was adopted in Ref.~\cite{Cheng:2013eu}, where the scale-dependent mass
anomalous dimension\footnote{Formally, the quark mass anomalous dimension is written as
$\gamma_m(g_R(\mu))$. Since we will be extracting $\gamma_m$
at different threshold eigenvalue parameters $M_R$, we will adopt the notation $\gamma_m(M_R)$ and speak
of the scale-dependent mass anomalous dimension, following the terminology of Ref.~\cite{Cheng:2013eu}.}
was extracted for an $SU(3)$ theory with 4 flavours of quarks.

Let us recall here the arguments relating the scaling of the mode number to the quark mass anomalous
dimension \cite{Cheng:2013eu}, paying special attention to the differences between IR-conformal and
chirally broken systems.

\subsubsection*{Spectral density in perturbation theory}
It can be shown \cite{Cheng:2013bca} in one-loop PT that in systems with asymptotic
freedom, the following relation holds:
\begin{equation}
\label{eq:pt}
\rho(\lambda,g_R^2)=C\lambda^{\frac{4}{1+\gamma_m(g_R^2)}-1}, 
\end{equation} 
where $C$ is a normalization constant, $g_R$ is the renormalized coupling and the
scheme-independent one-loop quark mass anomalous dimension $\gamma_m(g_R^2)$ depends on the gauge
group and the fermion representation. 
Thus, the above relation holds, in principle, both in chirally-broken and IR-conformal systems
that are asymptotically free.

Note that, in principle, additional contributions to the spectral density can be present in general,
e.g. terms growing with lower powers of $M$. 
However, such terms would not be visible in PT, i.e. they would modify the above
one-loop expression by adding a term like $C'\lambda^{\frac{3}{1+\gamma_m(g_R^2)}-1}$.
The absence of such term at one-loop implies that the corresponding term in the mode number can only
appear non-perturbatively.
However, at high enough $M$, the importance of such terms will become negligible. 
Hence, it will not affect the hypothesis for numerical evaluation that at high enough $M$, the method
should allow for the extraction of the mass anomalous dimension.
It should also be mentioned that terms of such kind could show up also in the extraction of
the chiral condensate from the mode number \cite{Giusti:2008vb,Cichy:2013gja,Engel:2013rwa}.
It was shown in the analysis of Ref.~\cite{Cichy:2013gja} that such potential effects are
numerically small, if $M$ is small enough.
An analogous thing has to hold on the other end of the scale of $M$ -- for large enough $M$, any
lower order terms (with lower powers of $M$) have to be unimportant.
Hence, in the following, we will not consider such terms.

\subsubsection*{Renormalization group scaling of the quark mass and Dirac operator eigenvalues}
Let us now consider the renormalization group scaling of the eigenvalues of the Dirac operator and of the
quark mass.
We assume that the quark mass is multiplicatively renormalizable:
\begin{equation}
 m_R(g_0,\mu)=Z_m(g_0,\mu)m
\end{equation}
where $g_0$ is the bare coupling and the mass renormalization constant $Z_m(g_0,\mu)$ is mass-independent
(i.e. defined in a mass-independent renormalization scheme).
Let us start with the definition of the quark mass anomalous dimension:
\begin{equation}
\label{eq:def-gamma}
 \gamma_m(g_R(\mu))=-\frac{d\ln m_R(g_0,\mu)}{d\ln\mu}\Big|_{g_0=g_0(g_R(\mu))}
\end{equation} 
where $g_R(\mu)$ is the renormalized coupling and $g_0$ on the right-hand
side is such that it corresponds to the renormalized coupling $g_R(\mu)$ on the left-hand side (this will
be implied in the following formulae).
For two chosen scales $\mu_1$ and $\mu_2$, this equation can be rewritten as:
\begin{equation}
\label{eq:mR1}
 m_R(g_0,\mu_2)=m_R(g_0,\mu_1)\exp\left(-\int_{\ln\mu_1}^{\ln\mu_2}\gamma_m(g_R(\mu))\,d\ln\mu\right).
\end{equation}
Close to a renormalization group (RG) fixed point, $\gamma_m(g_R(\mu))$ depends very mildly on $\mu$,
i.e. $g_R(\mu_1)\approx g_R(\mu_2)$, and the above equation becomes:
\begin{equation}
\label{eq:mR2}
 m_R(g_0,\mu_2)=m_R(g_0,\mu_1)\left(\frac{\mu_2}{\mu_1}\right)^{-\gamma_m(g_R(\mu_1))}.
\end{equation}

One can now consider an RG transformation with a scale factor $b$, i.e. the scale
transforms as: $\mu\rightarrow\mu/b$.
The bare quark mass scales as: $m\rightarrow m/b$.
In the renormalized quark mass, there is an additional effect, coming from the running of the
renormalization constant with the scale.
Thus:
\begin{equation}
\label{eq:rgm}
 m_R(g_0,\mu_1) \rightarrow \frac{m_R(g_0,\mu_1)}{b} \frac{Z_m(g_0,\mu_2)}{Z_m(g_0,\mu_1)}=
\frac{m_R(g_0,\mu_1)}{b^{1+\gamma_m(g_R(\mu_1))}},
\end{equation} 
where the equality comes from Eq.~\eqref{eq:mR2}.
The volume transforms as: $V\rightarrow b^4 V$.
In analogy to the quark mass anomalous dimension $\gamma_m$, one can also define the anomalous dimension
of the threshold eigenvalue $M$ via Eq.~\eqref{eq:def-gamma} with the replacement $m\rightarrow M$.
By arguments similar to the above, one can then show the scaling relation under an RG transformation to
be: $M\rightarrow M/b^{1+\gamma_M}$.
Since the threshold $M$ and the twisted quark mass $m$ (we anticipate the use of twisted mass
fermions in the following) renormalize with the same renormalization
constant \cite{Giusti:2008vb}, they have to run with the same anomalous dimension,
i.e. $\gamma_M(g_R)\equiv\gamma_m(g_R)$.
In the following, we will use the notation $\gamma_M(M)\equiv\gamma_M(g_R(M))$, since $M$ plays the role
of the renormalization scale (see below).

Having shown the above scaling properties in the vicinity of an RG fixed point, we can now relate
$\gamma_M(M)$ to the scaling of the spectrum of $D^\dagger D$.
In the free theory $\nu(M)\propto VM^4$, while
interactions modify this scaling behaviour
to $\nu(M)\propto VM^\alpha$, where $\alpha$ is close to 4 if the eigenvalues are large, i.e.
correspond to the ultraviolet.
The renormalization group invariance of the mode number, proved in Ref.~\cite{Giusti:2008vb},
implies:
$VM^\alpha=b^4 V \left(M/b^{1+\gamma_m(M)}\right)^\alpha$ and
hence $\alpha=4/\left(1+\gamma_m(M)\right)$.
Thus, $\nu(M)\propto M^{4/\left(1+\gamma_m(M)\right)}$ and the scale-dependent anomalous dimension
$\gamma_m(M)$ parametrizes deviations of the mode number scaling from the free-field theory value of
4, in accordance with the PT formula \eqref{eq:pt}.
In this way, we have established the relation of the anomalous dimension $\gamma_m(M)$ to the scaling of
the mode number.
At this point, it is important to emphasize that this relation is not universally valid and is subject
to several conditions restricting its range of applicability.
It is therefore essential to discuss these limitations.

\begin{figure}
\begin{center}
\includegraphics
[width=0.6\textwidth,angle=270]
{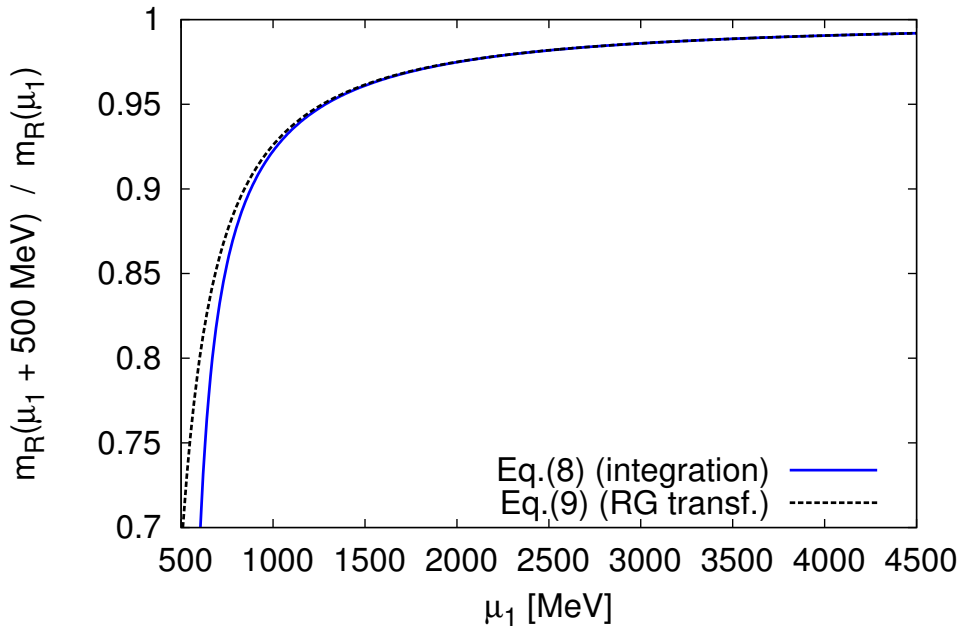}
\caption{The ratio of $\MSb$ renormalized quark masses ($N_f=2$ QCD) at scales $\mu_1$ and $\mu_2$
such that $\mu_2=\mu_1+500$ MeV. The blue solid line is the result from Eq.~\eqref{eq:mR1} (exact
integration using 4-loop $\gamma_m(g_R(\mu))$) and the black dashed line from Eq.~\eqref{eq:mR2} (RG
transformation valid close to the (ultraviolet) RG fixed point), using 4-loop
$\gamma_m(g_R(\frac{\mu_1+\mu_2}{2})$).}
\label{fig:compare}
\end{center}
\end{figure}

\subsubsection*{Limitations of applicability of the approach}
First of all, let us reconsider Eq.~\eqref{eq:mR1}. In general, i.e. if the implicit scale dependence of
$\gamma_m(g_R)$ is non-negligible, Eq.~\eqref{eq:mR1} can not be written as \eqref{eq:mR2} -- it becomes:
\begin{equation}
\label{eq:mR3}
 m_R(g_0,\mu_2)=m_R(g_0,\mu_1)\exp\left(\int_{g_R(\mu_1)}^{g_R(\mu_2)}\,dg_R\frac{\gamma_m(g_R)}{
\beta(g_R)}\right),
\end{equation}
with the running of the coupling given by the $\beta$-function.
In this general case, Eq.~\eqref{eq:rgm} is modified such that the relation between the scaling exponent
$\alpha$ of the mode number and the mass anomalous dimension $\gamma_m$ receives corrections that make
the functional form of this relation more complicated (it then includes the integral of
Eq.~\eqref{eq:mR1} instead of the factor $b^{-\gamma_m}$).
If so, one would either need some perturbative input to evaluate this integral or the values of the mode
number at all values of $M$ in the analyzed interval (and not only at selected values of $M$ separated by
$\mathcal{O}(100)$ MeV, as will be done in practice -- see Secs.~\ref{sec:str1} and \ref{sec:procedure}),
increasing by far the numerical cost (see also Sec.~\ref{sec:specproj} for a comment on this cost in
the present context).
It is therefore of utmost importance to check the relevance of this limitation.
As we will show now, this relevance is minor in practice (in the case of QCD), if one
restricts the numerical analysis to intervals in the threshold eigenvalue $M$ of a few hundred MeV and if
the considered values of $M$ are large enough, i.e. close enough to the RG ultraviolet fixed point.
This is demonstrated in Fig.~\ref{fig:compare}, which shows the ratio of renormalized quark masses
$m_R(\mu_2)/m_R(\mu_1)$, in the $\MSb$ scheme at 4-loops, in QCD with 2 flavours of dynamical quarks, for
scales satisfying $\mu_2=\mu_1+500$ MeV.
The interval of 500 MeV was chosen to match our intervals for extracting $\gamma_m$ from the lattice. 
The blue solid line shows the result from the application of Eq.~\eqref{eq:mR1} (or, equivalently,
Eq.~\eqref{eq:mR3}), which takes into account the $g_R$-dependence of $\gamma_m$ and hence its implicit
$\mu$-dependence. In this sense, this result is exact, in contrast to the result from Eq.~\eqref{eq:mR2}
(black dashed line in Fig.~\ref{fig:compare}), which is, in principle, valid only in the vicinity of a
RG fixed point (ultraviolet in this case). To compute the ratio $m_R(\mu_2)/m_R(\mu_1)$ according to
Eq.~\eqref{eq:mR2}, we have inserted $\gamma_m$ evaluated in the middle of the interval from
$\mu_1$ to $\mu_2$, i.e. $\gamma_m(g_R(\frac{\mu_1+\mu_2}{2}))$.
Inspection of Fig.~\ref{fig:compare} shows that the difference between both considered results
is minor if $\mu_1$ is larger than approx. 1 GeV. Numerically, this difference amounts to 0.001\% at 4
GeV (i.e. for $\mu_1=4$ GeV and $\mu_2=4.5$ GeV), 0.016\% at 2 GeV, 0.05\% at 1.5 GeV and 0.38\% at 1
GeV. Only below $\mu_1=1$ GeV, the effects start to be larger than our statistical errors.
However, at such low energies, non-perturbative effects are beginning to be essential and comparison
to PT does not make much sense.
The chosen difference $\mu_2-\mu_1=500$ MeV is a good
compromise between availability of lattice data and the discussed effect, which is below 0.5\%
for scales above 1 GeV. However, even choosing $\mu_2-\mu_1=2$ GeV would lead to effects below 0.5\% for
scales above approx. 1.7 GeV. This results from the following numerical observation --
the contribution to the integral in Eq.~\eqref{eq:mR1} of the intervals
$[\mu_1,(\mu_1+\mu_2)/2]$ and $[(\mu_1+\mu_2)/2,\mu_2]$ balances out as if $g_R$ was
constant and equal to the value in the middle of the interval and could hence be taken out of the
integral.
In this way, the use of Eq.~\eqref{eq:mR2} to relate the quark mass anomalous dimension to the mode
number, instead of the exact Eq.~\eqref{eq:mR1} does not constitute a problem at relevant scales.
Note that this is to some extent surprising, since naively one could expect that Eq.~\eqref{eq:mR2} is
valid only in the immediate vicinity of the fixed point, e.g. at scales above 5 or 10 GeV, where no
contact to the lattice would be possible at presently simulated lattice spacings.

The second limitation that we would like to shortly discuss is related to the following issue.
Given the
equation for the massive Dirac operator $D_m^\dagger D_m=D^\dagger D+m^2$ (where $D$
is the massless operator), it is plausible to expect that, at least in some energy range,
the eigenvalues of the Hermitian Dirac operator scale in the same way as the quark mass $m$ or the
threshold mass $M$.
However, one has to keep in mind that such scaling is not universally valid for all
individual eigenvalues.
For instance, the behaviour of eigenvalues near the origin gives rise to the
chiral condensate in chirally broken systems according to the Banks-Casher relation.
Moreover, some eigenvalues are unphysical, e.g. related to the doubler modes (that decouple only
strictly in the continuum limit), and the concept of renormalization is ill-defined for them.

\subsubsection*{Importance of a numerical check of the approach}
Summarizing, the domain of applicability of the argument relating the dependence of the mode
number on the threshold mass $M$ and the anomalous dimension $\gamma_M$ has to be checked
\emph{numerically} and is not expected to be universally valid, e.g. the behaviour of $\nu(M)$ for very
small $M$ gives information
about the chiral condensate~\cite{Giusti:2008vb} and not about the anomalous dimension $\gamma_m$.
A priori, therefore, one expects the results obtained from this analysis to be valid for rather large
values of $M$, i.e. such that the properties of the system are governed by perturbative effects.
Hence, it is a priori unclear whether it is possible to address these issues with a lattice calculation
with presently available lattice spacings (with inverse cut-offs of the order of a few GeV).

The aim of the present paper is to investigate in practice the above mentioned problems, by looking at
the Dirac operator spectrum in lattice QCD with $N_f=2$ flavours of dynamical twisted
mass quarks.
In particular, we want to check if it is possible to recover values of the mass anomalous dimension
predicted by PT for some range of energies where contact can be established between
the latter and the lattice theory. This requires either an appropriate renormalization of the
threshold parameter $M$, which will be the subject of the next subsection, or matching to PT.

\subsection{Renormalization and quark mass anomalous dimension in perturbation theory}
\label{sec:renorm}

The mass scale (threshold for eigenvalues of $D^\dagger D$) $M$ is a bare scale and has to be
renormalized in order to make contact with PT.
Let us start with an analogy with the computation of the renormalized chiral condensate from the
Dirac operator spectrum \cite{Giusti:2008vb,Cichy:2013gja}.
The bare chiral condensate $\Sigma$ is extracted from the slope of the mode number vs. $M$ dependence.
Since the product $M\Sigma$ is renormalization group invariant, to calculate  $\Sigma^{\MSb,\mu}$, the
renormalized condensate in the $\MSb$ scheme, at some
scale $\mu$, the threshold $M$ has to be renormalized with the renormalization constant
$Z_P$, in the $\MSb$ scheme, at the scale $\mu$, i.e. $M$ is renormalized according to
$M_R=(Z_P^{\MSb,\mu})^{-1}\,M$:
\begin{equation}
\Sigma^{\MSb,\mu}=Z_P^{\MSb,\mu}\Sigma\sim
Z_P^{\MSb,\mu}\,\frac{\partial\nu(M,m)}{\partial M}=\frac{\partial\nu_R(M_R,m_R)}{\partial M_R},
\end{equation}
which boils down to an extraction of the slope of the renormalized mode number vs. renormalized mass
scale $M_R$.

In the present case, we need to renormalize the threshold scale $M$ in an analogous way:
\begin{equation}
\label{eq:ren}
 M_R=Z_P^{-1}(\mu\!=\!M_R)\,M,
\end{equation}
i.e. each value of $M$ has to be renormalized with a separate value of $Z_P^{-1}$, computed at the
scale $\mu=M_R$ and expressed in the $\MSb$ scheme, since we want comparisons with mass anomalous
dimension defined in this scheme and
$M\Sigma=\left(Z_P^{-1}(\mu\!=\!M_R)\,M\right)\left(Z_P(\mu\!=\!M_R)\,\Sigma\right) =M_R\Sigma_R$ is
renormalization group invariant.
This renormalization condition will become a basis for our analysis strategy 1.

We remind here the expression for the quark mass anomalous dimension $\gamma_m$ in perturbation
theory \cite{Chetyrkin:1997dh,Vermaseren:1997fq}:
\begin{equation}
\label{eq:gamma_def}
-\frac{d\,\ln\,m_R(\mu)}{d\,\ln\,\mu^2}
 \equiv \gamma_m(a_s(\mu)) \equiv
\sum_{i\geq0}\gamma_{{i}}
a_s(\mu)^{i+1},
\end{equation}
where 
$a_s(\mu) = \alpha_s(\mu)/\pi= g_R(\mu)^2/(4\pi^2)$
and $\gamma_i$ are known coefficients for $i=0,\ldots,3$.
In  numerical form, $\gamma_m$ for $N_f$ flavours reads in the $\MSb$ scheme:
\begin{eqnarray}
\label{eq:gamma}
\gamma_m =  &-& a_s - a_s^2  (4.20833 - 0.138889 N_f)
-a_s^3  (19.5156 - 2.28412 N_f - 0.0270062 N_f^2 )  
\nonumber\\ 
&-&
a_s^4  (98.9434 - 19.1075 N_f + 0.276163 N_f^2  + 0.00579322 N_f^3 ),
\end{eqnarray}
up to 4 loops \footnote{Note that in Eq.~\eqref{eq:gamma_def}, the anomalous dimension is defined as
$\gamma_m=-\frac{d\,\ln\,m_R(\mu)}{d\,\ln\,\mu^2}$, while Eq.~\eqref{eq:spectral} and the following
equations of Sec.~\ref{sec:spectral} define it as $\gamma_m=-\frac{d\,\ln\,m_R(\mu)}{d\,\ln\,\mu}$
(convention of e.g. Ref.~\cite{Ryttov:2007cx}), i.e. $\gamma_m$ extracted from these
formulae has to be divided by 2 to compare to Eq.~\eqref{eq:gamma}. In the following, we do so, i.e.
we define the anomalous dimension in accordance with
Refs.~\cite{Chetyrkin:1997dh,Vermaseren:1997fq}.}.
The running of the coupling depends on the $\Lambda$-pearameter of the theory.
Hence, the continuum extrapolated lattice data for the scale dependence of $\gamma_m$, or equivalently
$m_R$, allow to extract the $\Lambda$-parameter, denoted by $\Lambda^{(2)}_{\MSb}$, where the
superscript stands for 2 flavours and the subscript for the renormalization scheme.

\subsection{Spectral projector method of computing the mode number}
\label{sec:specproj}
The mode number $\nu(M)$, i.e. the number of eigenvectors of the massive Hermitian Dirac operator
(with quark mass $m$) $D_m^\dagger D_m$ with eigenvalue magnitude below the threshold value of $M^2$,
can be computed essentially with two methods. 
First, one can explicitly compute some given amount of $n$ eigenvectors for each gauge field
configuration.
This gives the mode number $\nu(M)$ for each value of $M$ below the average eigenvalue
corresponding to the $n$-th eigenvector.
However, for our present goals, this method is too expensive in terms of computing time, since we
want to reach threshold values of $M$ corresponding to mode numbers of $\mathcal{O}(10^5)$.
This makes the other available method, the method of spectral projectors described in
Ref.~\cite{Giusti:2008vb}, very attractive.
Its main advantage is its good scaling with the volume -- the cost of the computation scales with
$V$, instead of $V^2$ as in the case of an explicit computation.
Note that the usage of the spectral projector method (i.e. computation of the mode number only for
$\approx20$ selected values of $M$ spanning a range of a few GeV) is allowed for present purposes because
the relation between the scaling exponent of the mode number and $\gamma_m$ can be derived using
Eq.~\eqref{eq:mR2} and not explicitly using the integral in Eq.~\eqref{eq:mR1} -- else one would need the
full continuous dependence of the mode number on the threshold eigenvalue $M$ (as discussed in
Sec.~\ref{sec:spectral}).
In this way, one could reach only mode numbers of at most $\mathcal{O}(10^3)$ (and even this with a
higher computational effort than using the spectral projector method) and hence values of $M$ much
smaller than the inverse lattice spacing. 

For the spectral projector evaluation of the mode number, we use the implementation in the tmLQCD code
\cite{Jansen:2009xp}.
This implementation was extensively tested and used in our chiral condensate computation
\cite{Cichy:2013gja}.

Here, we shortly describe the method of spectral projectors. We refer to the original work of
Ref.~\cite{Giusti:2008vb} for a more complete account.
Let us define the orthogonal projector $\mathbbm{P}_M$ to the subspace of fermion fields spanned by
the lowest lying eigenmodes of the massive Hermitian Dirac operator $D_m^\dagger D_m$, with
eigenvalues below some threshold value $M^2$. 
The mode number $\nu(M)$ can be represented stochastically by:
\begin{equation}
 \nu(M)=\langle\textrm{Tr}\,\mathbbm{P}_M\rangle=\left\langle
\frac{1}{N}\sum_{j=1}^N (\eta_j,\mathbbm{P}_M\eta_j)\right\rangle,
\end{equation}
where $N$ pseudofermion fields $\eta_i$ are added to the theory.

The orthogonal projector $\mathbbm{P}_M$ can be approximated by a rational
function of $D_m^\dagger D_m$:
\begin{equation}
 \mathbbm{P}_M \approx h(\mathbbm{X})^4,\qquad
\mathbbm{X}=1-\frac{2M_*^2}{D_m^\dagger D_m+M_*^2},
\end{equation}
where the function:
\begin{equation}
 h(x)=\frac{1}{2}\left(1-xP(x^2)\right)
\end{equation} 
is an approximation to the step function $\theta(-x)$ in the range $-1\leq x\leq1$
and
$P(y)$ is in our case the Chebyshev polynomial of some adjustable degree $n_{\rm Chebyshev}$ that
minimizes the
deviation:
\begin{equation}
 \delta=\max_{\epsilon\leq y\leq 1}|1-\sqrt{y}P(y)|
\end{equation} 
for some $\epsilon>0$.
Computing the approximation to the
spectral projector $\mathbbm{P}_M$ requires solving the following equation an
appropriate number of times:
\begin{equation}
 (D_m^\dagger D_m + M_*^2)\psi=\eta
\end{equation} 
for a given source field $\eta$.
The parameter $M_*$ is related to the spectral threshold value $M$ and the ratio of $M/M_*$ depends on
the details of the approximation to the projector.
For our choice, $M/M_*\approx0.96334$ (as shown in Ref.~\cite{Giusti:2008vb}).


\section{Analysis strategy}
\label{sec:fit}
\subsection{Strategy 1 -- using $Z_P$ as an input}
\label{sec:str1}
The basic relation \eqref{eq:nu3} can be used to extract the scale dependence of the quark
mass anomalous dimension.
We insert the renormalization condition \eqref{eq:ren} and rewrite this relation as:
\begin{equation}
\label{eq:nu4}
\nu_R(M_R)=\nu_0(m_R)+\hat\rho\mu^{\frac{4\gamma_m(M)}{1+\gamma_m(M)}} \,
Z_P^{-\frac{4}{1+\gamma_m(M)}}
M^{\frac{4}{1+\gamma_m(M)}}.
\end{equation} 
Since we will work far away from the scale of spontaneous chiral symmetry breaking
$\Lambda_\chi$, we will ignore the term $\nu_0(m_R)$, having checked that it indeed does not
influence the extracted values of the mass anomalous dimension if $M_R\gtrsim1.5$ GeV (see below).
The contribution of this term can also be estimated from the spectral density at the origin $\rho(0)$,
which gives the value of the chiral condensate according to the Banks-Casher relation
\cite{Banks:1979yr}.
In the low-energy regime $\rho(\lambda)$ is approximately constant and equal to its value at the origin
($\Sigma/\pi$ from the Banks-Casher relation) -- $\nu(M)$ grows linearly with $M$ and
its slope determines the chiral condensate and hence it gives a rough estimate of the contribution of
the term $\nu_0(m)$ to the total mode number (the dependence on the quark mass $m$ is implicit in the
value of the chiral condensate):
\begin{equation}
\label{eq:nu0}
\nu_0(m)=2V\int_0^{\Lambda_\chi}d\lambda\,\rho(\lambda)\approx
\frac{2}{\pi}\Sigma\Lambda_\chi V.
\end{equation} 
The interval where the effects of spontaneous chiral symmetry breaking are important is assumed to lie
between the origin and some scale denoted by $\Lambda_\chi$. In the numerical part of this work, we will
consider several values of $\Lambda_\chi$ to check the robustness of the results with respect to effects
of spontaneous chiral symmetry breaking, by estimating the contribution of the term $\nu_0(m)$ to the
total mode number (the value of the chiral condensate for this estimate can be taken from the low part of
$\nu(M)$ vs. $M$ dependence).
Note that the above equation is just the leading-order chiral perturbation theory expression for the
effective chiral condensate -- see e.g. Eq.~(4.2) in Ref.~\cite{Giusti:2008vb}, with the identification
$\Lambda=\Lambda_\chi$.

Considering only a range of $M$ such that the term $\nu_0(M)$ can be safely neglected, one can rewrite
Eq.~\eqref{eq:nu4} as:
\begin{equation}
\label{eq:nu5}
 \nu_R(M_R)=\nu(M)\approx A\,M^{\frac{4}{1+\gamma_m(M)}},
\end{equation} 
where we have used the fact that the mode number is renormalization group invariant, i.e.
$\nu_R(M_R)=\nu(M)$ \cite{Giusti:2008vb}.
The quantity $A$ is approximately constant if the bare scale $M$ varies only little.
Eq.~\eqref{eq:nu5} can be fitted \emph{locally} (in short intervals in $M$ , i.e. such that
$A$ and $\gamma_m$  can be considered constant) for various \emph{bare} scales $M$.
It has 2 fitting parameters: $A$ and $\gamma_m(M)$.
Using the renormalization condition \eqref{eq:ren}, the values of $\gamma_m(M)$ can be translated to
$\gamma_m(M_R)$.
Since we extract $\gamma_m(M_R)$ from lattice data, we expect that the obtained values of
$\gamma_m(M_R)$ are equal to $\gamma_m(a_s(M_R))$ of PT only up to cut-off effects:
\begin{equation}
\gamma_m(M_R,a)=\gamma_m(a_s(M_R))+\mathcal{O}(a),
\end{equation} 
where we symbolically write that the extracted values of $\gamma_m(M_R)$ depend on the lattice
spacing $a$.
We will use the twisted mass Dirac operator (see next section) to extract the mass anomalous
dimension.
The twisted mass Dirac operator gives automatic $\mathcal{O}(a)$-improvement in physical
($\mathcal{R}_5$-parity even) quantities \cite{Frezzotti:2003ni}.
In particular, the mode number of the Dirac operator is automatically $\mathcal{O}(a)$-improved
\cite{Cichy:2013egr}. 
However, this does not necessarily imply the improvement of $\gamma_m$.
Inspecting Eq.~\eqref{eq:nu3}, it can not be excluded that the mode number is $\mathcal{O}(a)$-improved
even if the factors that enter it have $\mathcal{O}(a)$ effects.
Hence, the $\mathcal{O}(a)$-improvement of $\gamma_m$ can not be concluded from the improvement of the
mode number.
For this reason, the continuum limit extrapolations in the numerical part will be performed under the
assumption that $\mathcal{O}(a)$ effects can be present.
With a rather good precision of the method, it can turn out \emph{a posteriori} that the coefficient of
the $\mathcal{O}(a)$-term in the continuum limit extrapolations of $\gamma_m$ is compatible with zero.
Actually, this will not be the case -- see Sec.~\ref{sec:cont} and Appendix \ref{sec:a2}.

We emphasize that only the continuum limit extrapolated values
$\gamma_m(M_R)=\lim_{a\rightarrow0}\gamma_m(M_R,a)$ can be compared to $\gamma_m(a_s(M_R))$,
provided that the lattice window exists for the contact of lattice simulations with
continuum PT (which can be written schematically as:
$\mathcal{O}(\Lambda_{\rm QCD})\ll M\ll a^{-1}$).
 
A further check of the method can be performed by rewriting equation for the $M_R$-dependence of
$\nu_R(M_R)$ \eqref{eq:nu3} at the renormalization scale $\mu=M_R\,$:
\begin{equation}
\label{eq:nu6}
\nu_R(M_R)=\hat\rho\mu^{\frac{4\gamma_m(M_R)}{1+\gamma_m(M_R)}} \,
M_R^{\frac{4}{1+\gamma_m(M_R)}}=\hat\rho M_R^4.
\end{equation} 
This equation implies that the renormalized mode number scales with the fourth power of the
renormalized threshold parameter $M_R$ for all values of the latter.
In practice, cut-off effects can lead to deviations from the above statement.
Therefore, we write Eq.~\eqref{eq:nu6} as:
\begin{equation}
\label{eq:nu7}
 \nu_R(M_R)\approx A'\,M_R^{\frac{4}{1+\Gamma_m(M_R)}},
\end{equation} 
with an ``anomalous dimension'' $\Gamma_m(M_R)$ which is purely a lattice artefact, i.e. its continuum
limit should be zero.
Hence, we will call it the ``artefact'' anomalous dimension.

\subsection{Strategy 2 - matching to perturbation theory}
\label{sec:str2}
An alternative method of analysis was proposed in Ref.~\cite{Cheng:2013eu}. It does not require the
knowledge of the renormalization constant $Z_P$ and instead matching to PT is
performed.
It consists in selecting one value of $\beta$ as the reference value (denoted by $\beta_{\rm ref}$)
and rescaling lattice eigenvalues for other bare couplings to express them in terms of a uniform scale
$a_{\rm ref}$, i.e. the lattice spacing corresponding to $\beta_{\rm ref}$.
The rescaling of eigenvalue $M_\beta$ at a given value of $\beta$ is as follows:
\begin{equation}
 M_\beta \rightarrow M_\beta \left(\frac{(r_0/a)_\beta}{(r_0/a)_{\rm
ref}}\right)^{1+\gamma_m(M_\beta)}.
\end{equation} 

Then, we extract the anomalous dimension $\gamma_m(M_{\rm ref})$ for all lattice spacings by
fitting Eq.~\eqref{eq:nu5} and perform continuum limit extrapolations at fixed values of $M_{\rm
ref}$ (i.e. with eigenvalues at all lattice spacings rescaled to correspond to the chosen reference
$\beta$).
Finally, we choose one value of $M_{\rm ref}$ for matching to PT, employing the
matching condition:
\begin{equation}
 \gamma_m(M_{\rm ref,matching})=\gamma_m(\mu_{\rm matching}).
\end{equation} 
In this way, we set the scale, i.e. we know that $M_{\rm ref,matching}$ corresponds to $\mu_{\rm
matching}$, which is known in physical units.
A non-trivial test of the approach is provided by comparing the scale dependence of the lattice
extracted anomalous dimension with PT prediction.
Since the latter is scheme-dependent and our procedure implicitly defines a renormalization scheme, we
can only compare to the universal one-loop PT expression.

\section{Lattice setup}
\label{sec:setup}
Our lattice setup consists of the tree-level Symanzik improved gauge action \cite{Weisz:1982zw} and
the Wilson twisted mass fermion action
\cite{Frezzotti:2000nk,Frezzotti:2003ni,Frezzotti:2004wz,Shindler:2007vp}.
The former reads: 
\begin{equation}
 S_G[U] = \frac{\beta}{3}\sum_x\Big( b_0 \sum_{\mu,\nu=1} \textrm{Re\,Tr} \big( 1 - P^{1\times
1}_{x;\mu,\nu}
\big) 
+ b_1 \sum_{\mu \ne \nu} \textrm{Re\,Tr}\big( 1 - P^{1 \times 2}_{x; \mu, \nu} \big) \Big),
\end{equation}
where $b_1 = -\frac{1}{12}$, $b_0=1-8b_1$, $\beta=6/g_0^2$, $g_0$ is the bare coupling, $P^{1\times
1}$, $P^{1\times 2}$ are the plaquette and rectangular Wilson loops, respectively.
The Wilson twisted mass fermion action is
given in the so-called twisted basis by:
\begin{equation}
 S_l[\psi, \bar{\psi}, U] = a^4 \sum_x \bar{\chi}(x) \big( D_W + m_{0} + i m \gamma_5 \tau_3
\big)\chi(x),
 \label{tm_light}
\end{equation}
where $m_{0}$ ($m$) is the bare untwisted (twisted) quark mass.
The renormalized light quark mass is given by $m_R=Z_P^{-1}m$.
The matrix $\tau^3$ acts in flavour space and $\chi=(u,\,d)^T$ is a
two-component vector in flavour space, related to the one in the physical basis by a chiral rotation.
The standard massless Wilson-Dirac operator $D_W$ is:
\begin{equation}
 D_W = \frac{1}{2} \big( \gamma_{\mu} (\nabla_{\mu} + \nabla^*_{\mu}) - a \nabla^*_{\mu} \nabla_{\mu}
\big),
\end{equation}
where $\nabla_{\mu}$ and $\nabla^*_{\mu}$ are the forward and backward covariant
derivatives.

One of the main advantages of the twisted mass formulation is that it allows for an automatic
$\mathcal{O}(a)$ improvement of physical observables, provided the hopping parameter $\kappa = (8+2 a
m_0)^{-1}$, is tuned to maximal twist by setting it to its critical value, at which the PCAC quark
mass vanishes
\cite{Frezzotti:2000nk,Farchioni:2004ma,Farchioni:2004fs,Frezzotti:2005gi,Jansen:2005kk}.
However, the spectrum of the Dirac operator itself is not improved -- hence we expect the extracted
values of the quark mass anomalous dimension to be contaminated by $\mathcal{O}(a)$ discretization
effects.

\begin{table}
\begin{center}
  \caption{\label{tab:setup}Parameters of ETMC $N_f=2$ gauge ensembles
\cite{Boucaud:2007uk,Boucaud:2008xu,Baron:2009wt}: the inverse bare coupling $\beta$,
lattice size $(L/a)^3\times(T/a)$, bare twisted light quark mass in lattice units $am$,
$r_0/a$ \cite{Blossier:2010cr}, lattice spacing $a$ \cite{Baron:2009wt,Jansen:2011vv}, physical
extent
of the lattice $L$ in fm.}
  \begin{tabular}{ccccccc}
    Ensemble & $\beta$ & lattice & $am$ & $r_0/a$ & $a$ [fm] & $L$ [fm]\\
\hline
   B$40.16 $  & 3.90 & $16^3\times32$   & 0.004   & 5.35(4) & 0.0790(26) & 1.3 \\
  B$40.24 $  & 3.90 & $24^3\times48$  & 0.004 &  5.35(4) & 0.0790(26) &  1.9  \\
  C$30.20$ & 4.05 & $20^3\times40$  & 0.003 & 6.71(4) & 0.0630(20) & 1.3 \\
  D$20.24$ & 4.20 & $24^3\times 48$  & 0.002 & 8.36(6) & 0.05142(83) & 1.2 \\
  E$17.32$ & 4.35 & $32^3\times64$  & 0.00175 & 9.81(13) & 0.0420(17) & 1.3 \\
  \end{tabular}
\end{center}
\end{table}

\begin{table}
\begin{center}
  \caption{\label{tab:zp}
Renormalization constant $Z_P$ in the $\MSb$ scheme for the ETMC ensembles used in this work.
We give values of $Z_P$ at two scales -- $\mu_1=2$ GeV (given in
Refs.~\cite{Constantinou:2010gr,Alexandrou:2012mt,Cichy:2012is}) and another scale $\mu_2$ for which
no perturbative running to 2 GeV has been performed -- see text for more details.}
  \begin{tabular}{ccccc}
    $\beta$ & $\mu_1$ [GeV] & $Z_P^{\MSb,\mu_1}$ & $\mu_2$ [GeV] &
$Z_P^{\MSb,\mu_2}$\\
\hline
   3.90 & 2 & 0.437(7) & 2.5 & 0.461(7)\\
  4.05 &  2 & 0.477(6) & 3.1 & 0.524(7)\\
  4.20 & 2 & 0.501(13) & 3.8 & 0.573(15)\\
  4.35 & 2 & 0.503(6) & 1.8 & 0.487(15)\\
  \end{tabular}
\end{center}
\end{table}
  
Gauge field configurations that we have used for this work were generated by
the European Twisted Mass Collaboration (ETMC) with 
$N_f=2$ dynamical flavours of quarks \cite{Boucaud:2007uk,Boucaud:2008xu,Baron:2009wt}.
The details of lattice parameters considered for this work are shown in Tab. \ref{tab:setup}.
The linear extents of our lattices are relatively small, with $L\approx1.3$ fm.
However, we checked the size of finite volume effects by including a larger physical
volume for $\beta=3.9$, $am=0.004$, with $L/a=24$, i.e. a physical volume of around 1.9 fm.

To make comparisons to PT employing Strategy 1, we need values of the scale $M_R$ in
physical units.
To convert from bare $aM$ in lattice units to $M_R$ in MeV, we need the lattice spacing values
\cite{Baron:2009wt,Jansen:2011vv} (we take the uncertainty of the values reported in
Refs.~\cite{Baron:2009wt,Jansen:2011vv} as our systematic error) and the values of $Z_P$ in the
$\MSb$ scheme \cite{Constantinou:2010gr,Alexandrou:2012mt,Cichy:2012is}.
We give values of $Z_P$ at two scales -- one of them ($\mu_1$) being the conventional scale of
2 GeV (values as given in Refs.~\cite{Constantinou:2010gr,Alexandrou:2012mt,Cichy:2012is}).
Such values are obtained in the RI-MOM scheme \cite{Martinelli:1994ty} ($\beta=3.9$, 4.05, 4.2) at the
scale $1/a$ or in the X-space scheme \cite{Martinelli:1997zc,Gimenez:2004me} ($\beta=4.35$) at some
chosen scale $1/X_0$, then converted to the $\MSb$ scheme and perturbatively evolved to 2 GeV.
As such, they rely on the perturbative expansion of the quark mass anomalous dimension.
Since we want to compare our final results to the ones implied by this expansion, we do not want to
renormalize $M$ using $Z_P$ that has this expansion as an input.
Hence, we use $Z_P$ at the scale $\mu_2$, which is the scale $1/a$ or $1/X_0$ of the non-perturbative
renormalization scheme used to compute it.
The values in the RI-MOM scheme or in the X-space scheme are then only converted to the $\MSb$ scheme 
using the formulae derived in Ref.~\cite{Chetyrkin:1999pq} (from the RI-MOM scheme) or
Ref.~\cite{Chetyrkin:2010dx} (from the X-space scheme).
Thus, no perturbative running of $Z_P$ is performed and the final extracted values of $\gamma_m$ in
the continuum do not have continuum perturbative $\gamma_m$ as an input.
Moreover, such procedure allows for a comparison of the predicted running of $Z_P$, which can be
evolved according either to the perturbative expansion of $\gamma_m$ or according to the
non-perturbatively determined $\gamma_m$ (different for different values of $\beta$ and hence
contaminated by lattice artefacts).

\section{Results}
\label{sec:results}
\subsection{Procedure of $\gamma_m$ extraction}
\label{sec:procedure}
\begin{figure}
\begin{center}
\includegraphics
[width=0.6\textwidth,angle=270]
{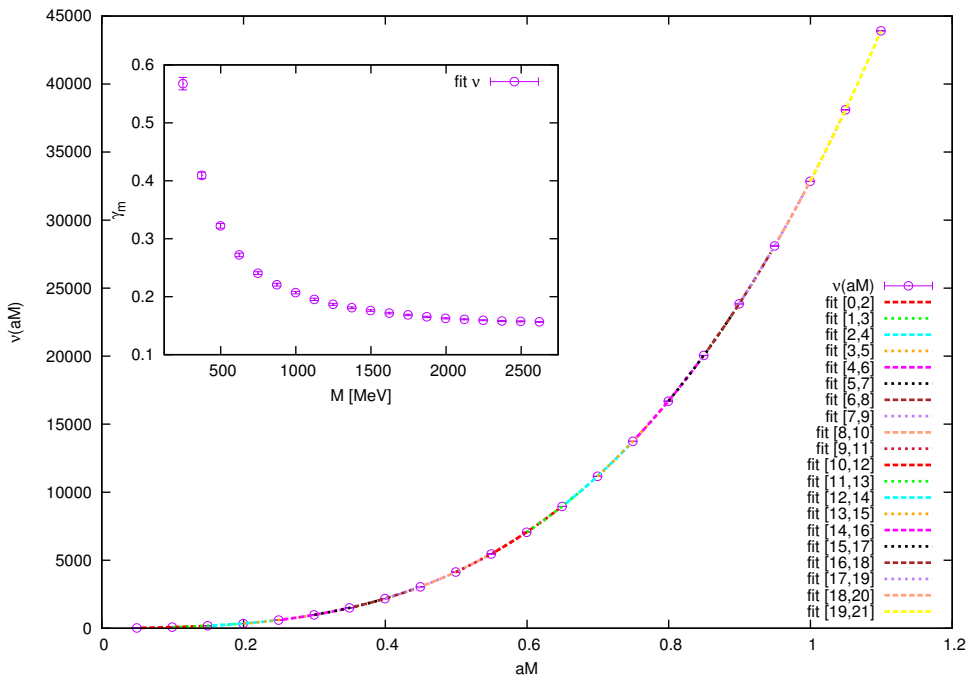}
\caption{Fits of Eq.~\eqref{eq:nu5} for the ensemble B40.16: $\beta=3.9$, $L/a=16$, $am=0.004$.}
\label{fig:example}
\end{center}
\end{figure}

\begin{figure}
\begin{center}
\includegraphics
[width=0.6\textwidth,angle=270]
{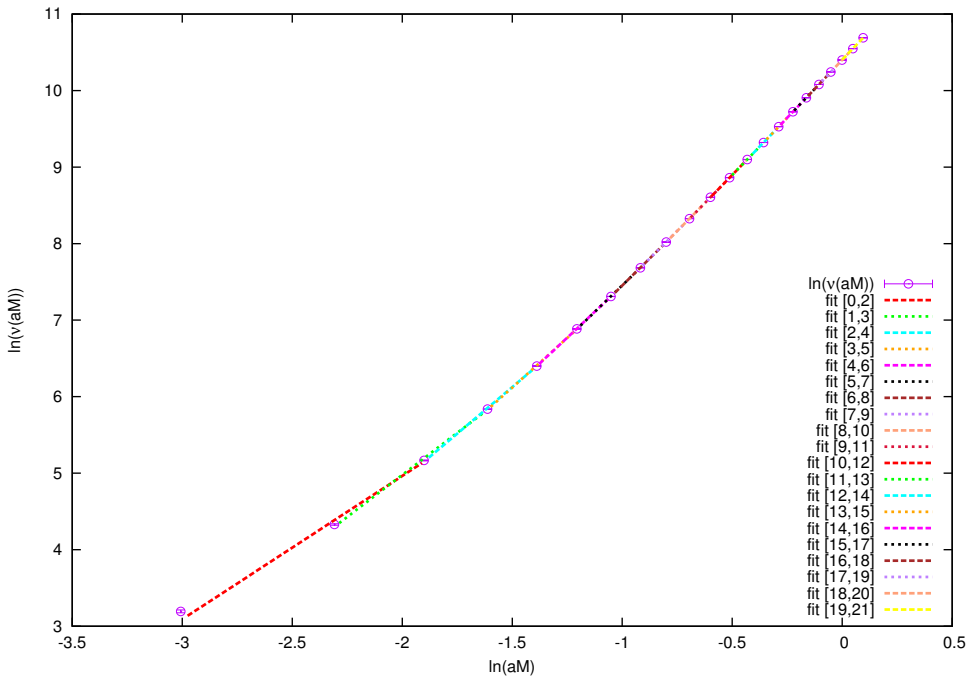}
\caption{Fits of Eq.~\eqref{eq:nu5} for the ensemble B40.16: $\beta=3.9$, $L/a=16$, $am=0.004$.
Log-log scale. For short enough intervals, $\ln\nu(aM)$ is a linear function of $\ln aM$.}
\label{fig:log}
\end{center}
\end{figure}

We start with an explicit example of our analysis strategy for ensemble B40.16, illustrated in
Figs.~\ref{fig:example} and ~\ref{fig:log}.
This part of the analysis is common to both strategies of analysis outlined in Sec.~\ref{sec:str1}
and \ref{sec:str2}. These strategies differ in the way lattice data are confronted with
continuum PT, which boils down to combining data at different lattice spacings
either by renormalizing eigenvalues and expressing them in physical units (Strategy 1) or by rescaling
them and matching to PT (Strategy 2).

The plots show 20 fits of Eq.~\eqref{eq:nu5}, corresponding to different fitting ranges.
The lattice data for the mode number are shown in the main plot of Fig.~\ref{fig:example} and in
Fig.~\ref{fig:log} in log-log scale.
For this ensemble, we used values of the bare threshold $aM$ in lattice units between 0.05 (approx.
125 MeV in physical units) and 1.10 (2750 MeV in physical units), with a step of 0.05 (125 MeV).
The first fit, labeled ``fit [0,2]'' includes the first 3 points and yields a value
$\gamma_m(M)=0.568(11)$, attributed to the middle value of the interval, i.e. $aM=0.1$ (250 MeV). This
value is then plotted in the inset of Fig.~\ref{fig:example}.
The following values plotted in the inset are from fits labeled ``fit [1,3]'' ($aM\approx 375$ MeV),
$\ldots$, ``fit [19,21]'' ($aM\approx 2625$ MeV).
In this way, we obtain the whole dependence of the anomalous dimension $\gamma_m(M)$ on the bare
scale $M$.

We remark here that the fits are stable with respect to:
\begin{itemize}
 \item including 3, 4 or 5 values of $\nu(M)$ in the fits,
 \item including the term $\nu_0(m)$ in the fits (see Eq.~\eqref{eq:nu4}).
\end{itemize}
The latter deserves a longer comment, as it regards the applicability of the method to theories with
spontaneous chiral symmetry breaking. 
In such theories, the spectral density of the Dirac operator does not go to zero near the origin, but
instead tends to a constant $\rho(0)$ and leads to a non-vanishing value of the chiral condensate.
Such constant value of the condensate produces a constant contribution to the mode number,
denoted by $\nu_0(m)$ and independent of $M$.
We have tried fits including different values of $\nu_0(m)$.
In particular, one can follow the discussion in Sec.~\ref{sec:fit} and use Eq.~\eqref{eq:nu0}.
We have calculated the condensate using data for the mode number vs. $M$, the latter
(unrenormalized) in the range between around 20 and 50 MeV \cite{Cichy:2013gja}. This range
corresponds
to a constant slope in the $M$-dependence of $\nu(M)$. Shortly above 50 MeV, one begins to see
deviations from linear behaviour, indicating an onset of a transitory region between a regime of
spontaneous chiral symmetry breaking and an intermediate regime in which we expect the scaling
relation of Eq.~\eqref{eq:nu1} might be valid.
We have tested the effect of 4 values of $\nu_0(m)$ on our fits,
corresponding to $\Lambda_\chi=125$, 250, 375 and 500 MeV.
For this ensemble, the bare condensate in lattice units $a^3\Sigma\approx0.0024$. 
This yields, respectively, $\nu_0(m)=10$, 20, 30 and 40 and
should be compared to the total mode number $\nu(M)$ at different values of $M$:
approx. 80 (at $M=250$ MeV), 350 (500 MeV), 2200 (1 GeV), 16700 (2 GeV).
The effect is sizable and can influence the extracted value of the anomalous dimension $\gamma_m(M)$
even around or slightly above $M=500$ MeV.
Indeed, at $M=500$ MeV, the extracted anomalous dimension is: 0.322(4), 0.297(4), 0.272(4), 0.246(4),
0.221(4) for $\nu_0(m)=0$, 10, 20, 30 and 40, respectively (the error is statistical only).
However, these apparent differences becomes much smaller after extrapolation to the continuum limit.
This will be shortly discussed again in Sec.~\ref{sec:cont} and in Appendix \ref{sec:nu0}.
For now, we anticipate the conclusion that the results in the continuum limit always agree for
different values of $\nu_0(m)\leq40$ if $M\gtrsim600$ MeV, even for the highest assumed value of
$\Lambda_\chi=500$ MeV.
We also remark that although the term $\nu_0(m)$ depends on the quark mass, in practice this
dependence is not relevant from the point of view of this analysis, as it varies the value of the
condensate (mass-dependent condensate defined in Ref.~\cite{Giusti:2008vb}) by at most 5\% when going
from quark mass $am=0.004$ towards the chiral limit, thus having negligible influence on
the value of $\nu_0(m)$.

\subsection{Finite volume effects}
\label{sec:fve}

\begin{figure}
\begin{center}
\includegraphics
[width=0.6\textwidth,angle=270]
{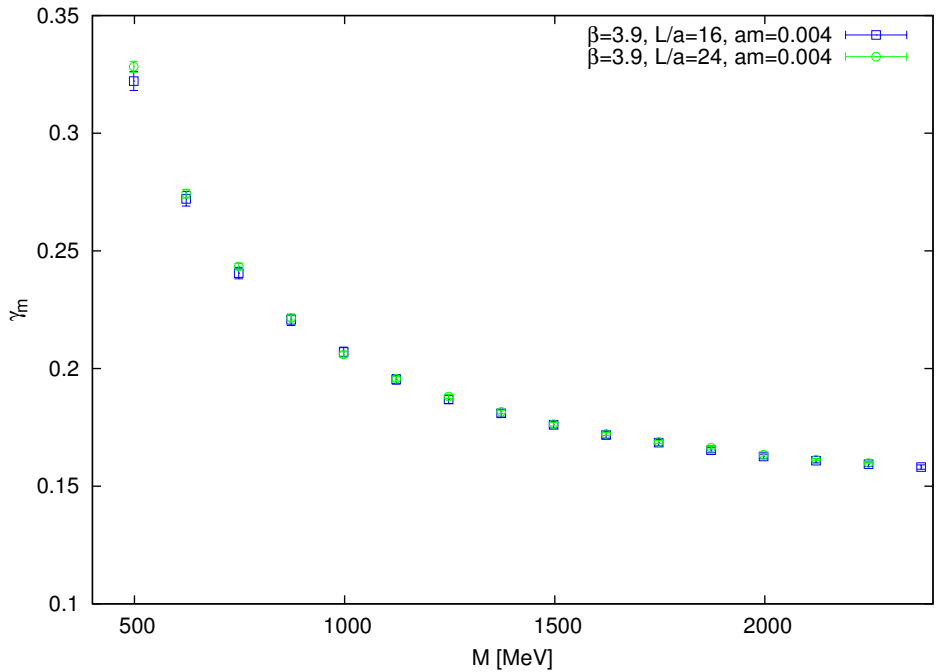}
\caption{Finite volume effects for the quark mass anomalous dimension, $\beta=3.9$,
$L/a=16$ and 24, $am=0.004$.}
\label{fig:fve}
\end{center}
\end{figure}

We have investigated finite volume effects by comparing the anomalous dimensions
$\gamma_m(M)$ calculated from ensembles B40.16 and B40.24, i.e. $\beta=3.9$, $am=0.004$ and $L/a=16$
or 24, which corresponds to $L\approx1.3$ or 1.9 fm.
It was found in Ref.~\cite{Cichy:2013gja} that finite size effects in the mode number density $\nu/V$
are small when one reaches a linear lattice extent of ca. 2 fm.
This is true if the threshold parameter $M\lesssim50$ MeV, i.e. in the range used for chiral
condensate extraction.
However, if $M$ is increased, finite size effects tend to decrease.
As argued in Ref.~\cite{Giusti:2008vb}, 
the difference between finite and infinite volume results for the chiral condensate and hence also
for the mode number density is of $\mathcal{O}(\exp(-M_\Lambda L/2))$, with
$M_\Lambda^2=2\Lambda\Sigma/F^2$, $\Lambda=\sqrt{M^2-\mu^2}$, $F$ is the pion decay constant in
the chiral limit.
The chiral condensate is extracted at $M\lesssim50$ MeV, while here we work with values of $M$ a
factor of 5-50 larger.
We can thus expect that we observe significant finite size effects only at the lower end of
considered values.

The results of comparison of B40.16 and B40.24 are shown in Fig.~\ref{fig:fve}.
We observe that the extracted values of $\gamma_m$ coincide for all values of $M$, ranging from
around 500 to 2500 MeV.
However, there is a systematic tendency towards discrepancy between the results from both ensembles
for small values of $M$, in accordance with theoretical expectations.
Finite volume effects above $M\gtrsim500$ MeV are small and the extracted values of the anomalous
dimension above this threshold can be considered to be infinite-volume results, thus justifying the
use of ensembles with $L\approx1.3$ fm.

\subsection{Non-perturbative running of $Z_P$}
In order to compare the extracted values of the quark mass anomalous dimension to the predictions of
PT, we have to take the continuum limit.
However, the values of $\gamma_m$ at non-zero lattice spacing can be used to perform non-perturbative
evolution of $Z_P$ (since $Z_P$ runs with the same anomalous dimension as the quark mass).

\begin{figure}
\begin{center}
\includegraphics
[width=0.34\textwidth,angle=270]
{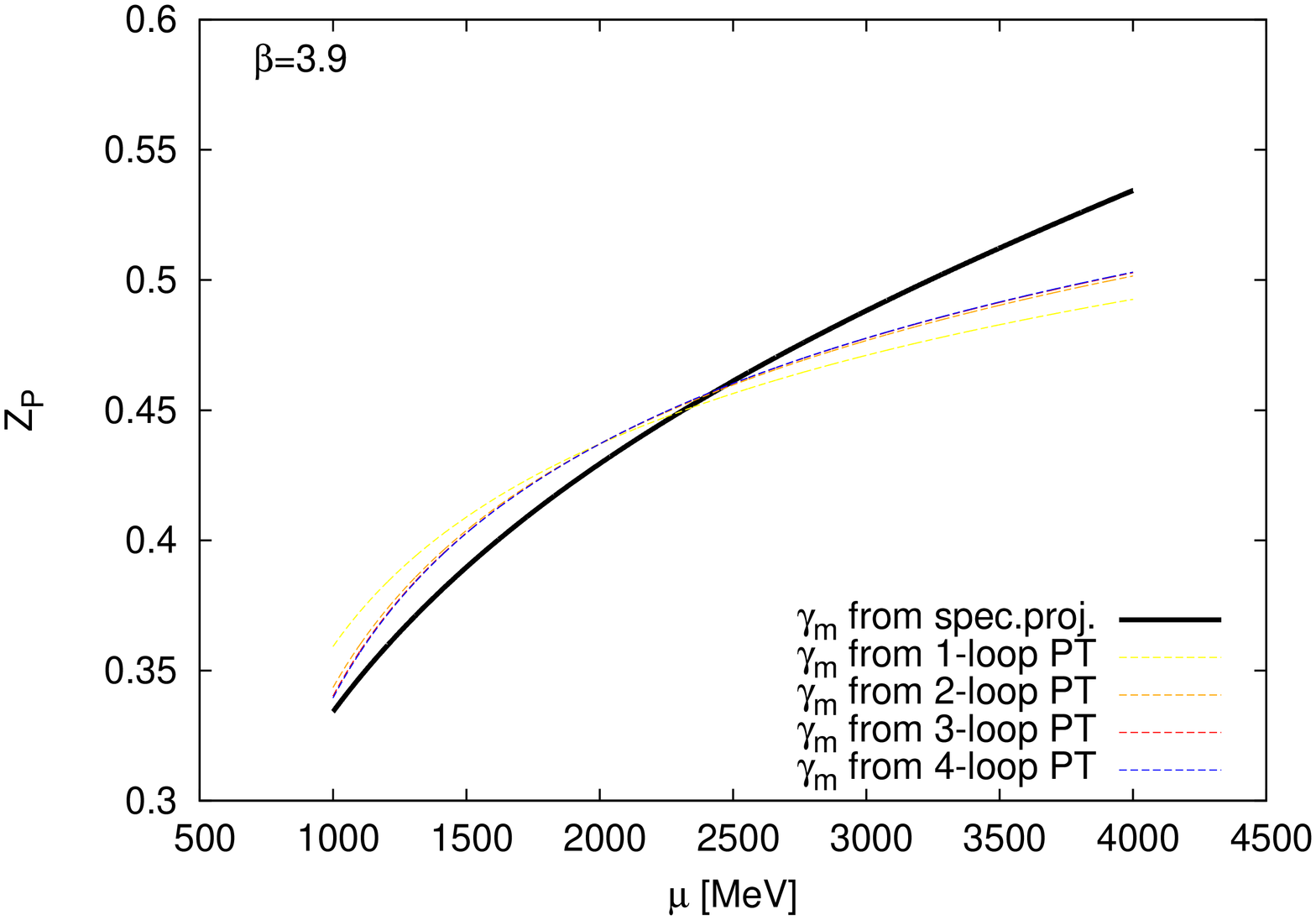}
\includegraphics
[width=0.34\textwidth,angle=270]
{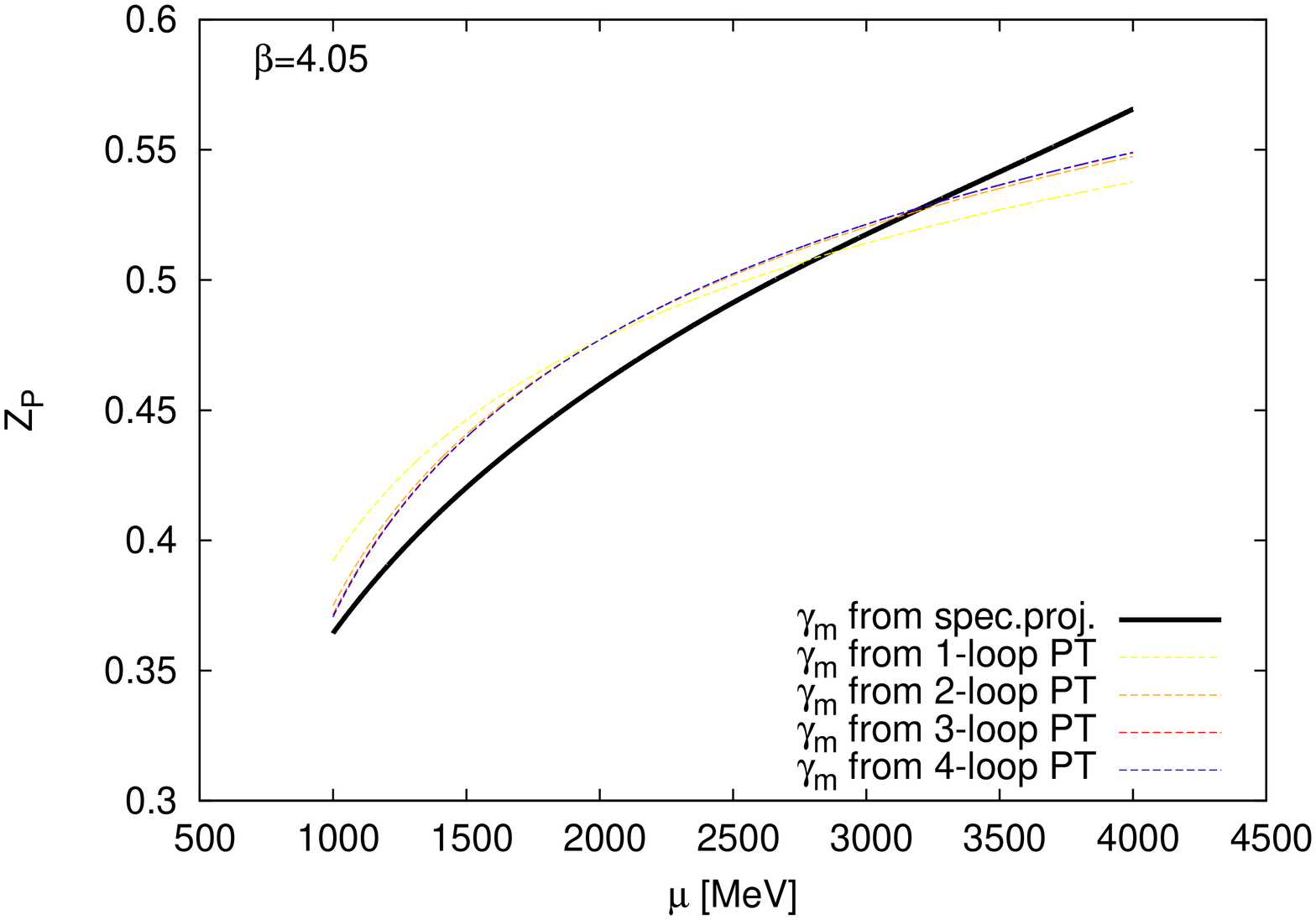}
\includegraphics
[width=0.34\textwidth,angle=270]
{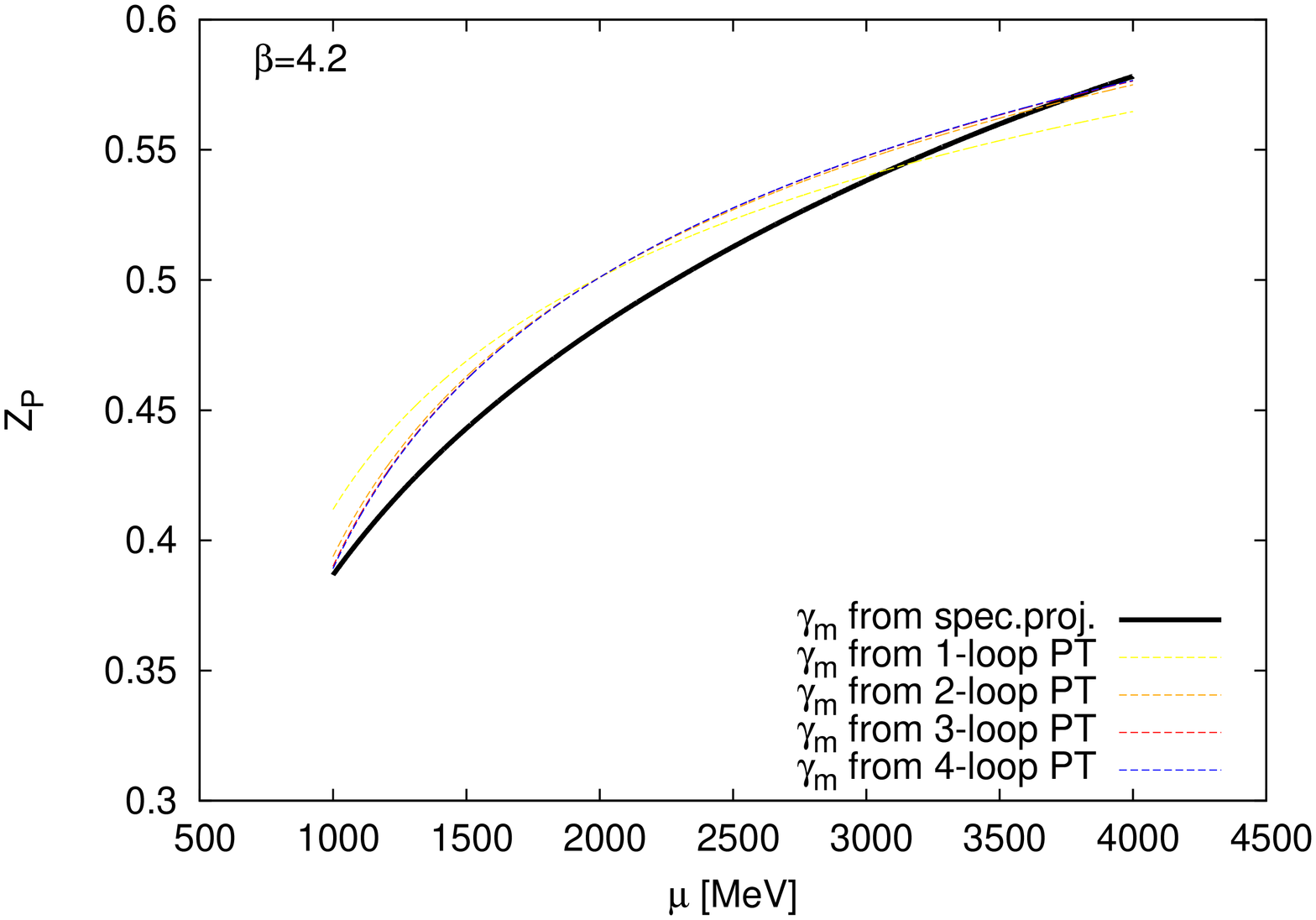}
\includegraphics
[width=0.34\textwidth,angle=270]
{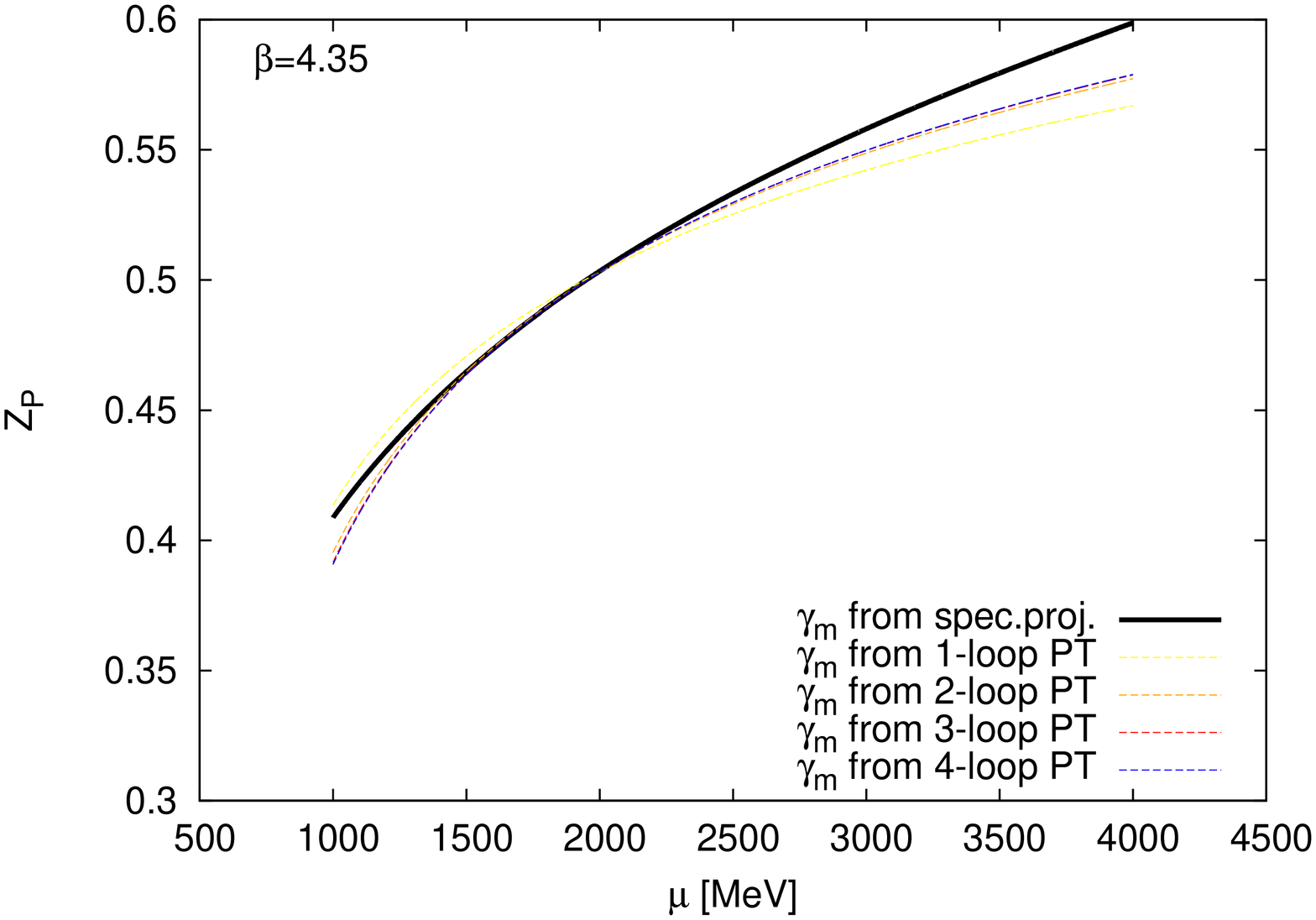}
\caption{Running of the renormalization constant $Z_P$ in the $\MSb$ scheme. Comparison of the
non-perturbative running (thick solid line) using quark mass anomalous dimension (converted
self-consistently to the $\MSb$ scheme) extracted from the mode number and the perturbative running
(1- to 4-loop, thin dashed lines) using the perturbative expansion of the anomalous dimension in the
$\MSb$ scheme, $\Lambda^{(2)}_{\MSb}=315$ MeV \cite{Jansen:2011vv}. The starting point of the evolution
of $Z_P$ is the scale $1/a$ ($\beta=3.9$,
4.05, 4.2) or $1/X_0\approx1.78$ GeV ($\beta=4.35$), i.e. at this scale $Z_P$ is the same for the
non-perturbative and 4-loop perturbative evolution. $Z_P$ for 1- to 4-loop perturbative evolution is
chosen equal at 2 GeV.}
\label{fig:ZP}
\end{center}
\end{figure}

The comparison of non-perturbative and perturbative evolution of $Z_P$ is shown in Fig.~\ref{fig:ZP}.
The former is extracted at the scale $1/a$ in the RI-MOM scheme or at some chosen scale $1/X_0$ in
the X-space scheme and converted to the $\MSb$ scheme at this scale.
Then, the lattice extracted quark mass anomalous dimension is used for the non-perturbative evolution
of $Z_P$.
Note that the evolution procedure is self-consistent, i.e. $\gamma_m$ is extracted at some bare scale
$M$ and this scale is renormalized using the values of $Z_P$ self-consistently evolved with this
$\gamma_m$ by numerically integrating the defining equation of $\gamma_m$:
\begin{equation}
 d\ln Z_P(\mu) = 2\gamma_m(\mu) d\ln\mu,
\end{equation} 
which leads to the relation:
\begin{equation}
\label{eq:ZPrunning}
 \ln Z_P(\mu_2)=\ln Z_P(\mu_1) + 2\gamma_m(M\!=\!Z_P(\mu_1)\mu_1)\ln\frac{\mu_2}{\mu_1},
\end{equation} 
which is used for small differences in $\mu_1$ and $\mu_2$ such that $\gamma_m$ can be considered
equal for $\mu_1$ and $\mu_2$.
The argument in the parentheses of $\gamma_m(M\!=\!Z_P(\mu_1)\mu_1)$ ensures that such evolution
procedure is self-consistent.

In general, the non-perturbative running of $Z_P$ is ``faster'', since lattice-extracted quark mass
anomalous dimensions are always larger at finite lattice spacing than their perturbative values (see
Fig.~\ref{fig:all_renorm}).
Since the difference with respect to the continuum values is an $\mathcal{O}(a)$ effect, the
non-perturbatively evolved $Z_P$ is
contaminated with additional $\mathcal{O}(a)$ effects from the extraction of the mass anomalous
dimension from the mode number.
However, we emphasize that the fact that we have the non-perturbative running of $Z_P$ will allow to
obtain in the end the scale dependence of $\gamma_m$ in the continuum and hence we will be able to
compare these values with the prediction of continuum PT.

\begin{figure}
\begin{center}
\includegraphics
[width=0.6\textwidth,angle=270]
{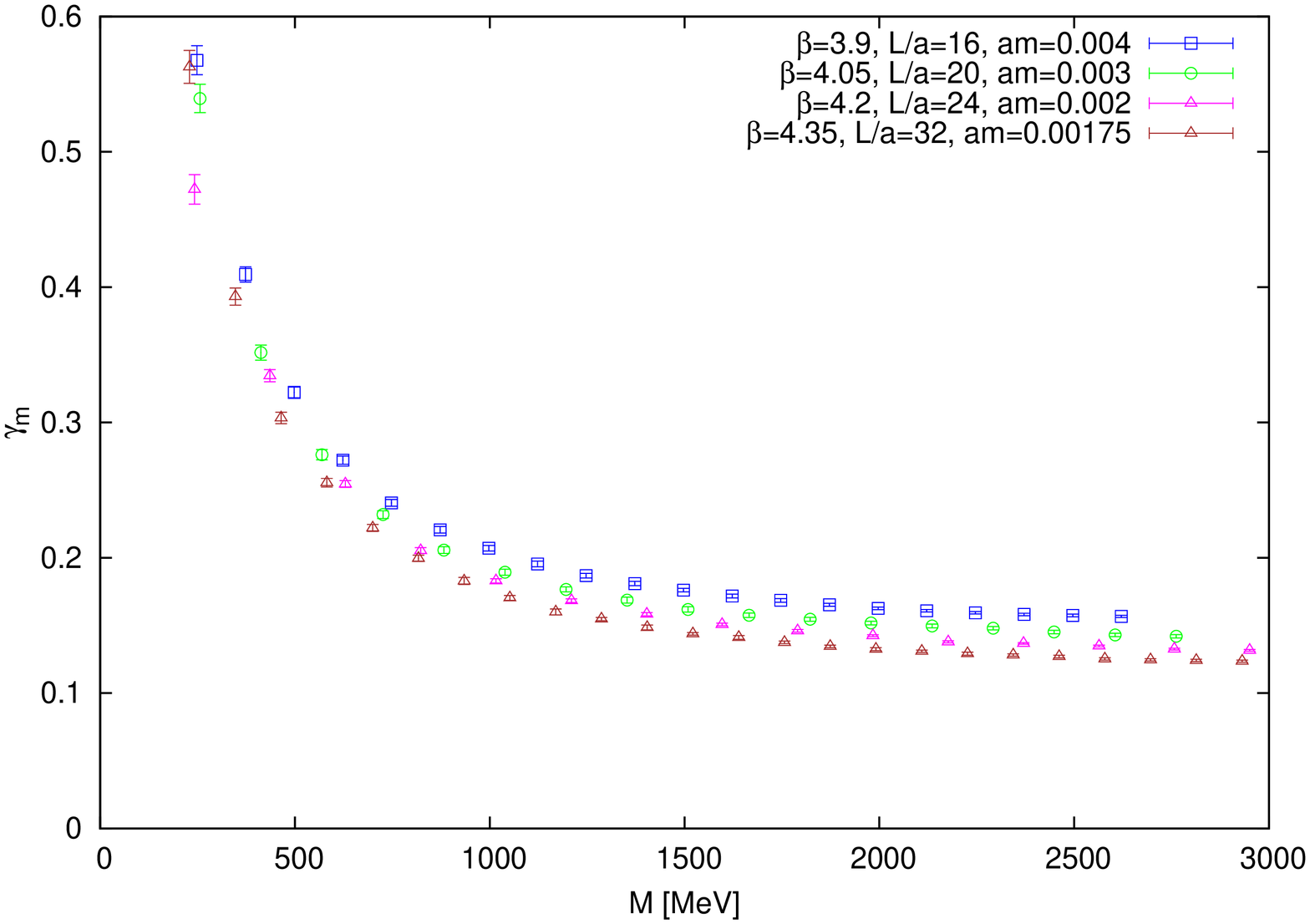}
\caption{Results for all 4 lattice spacings. The spectral threshold $M$ is unrenormalized.}
\label{fig:all_unrenorm}
\end{center}
\end{figure}

\subsection{Continuum limit -- Strategy 1}
\label{sec:cont}
We have repeated the procedure described in Sec.~\ref{sec:procedure} to extract the anomalous
dimension for four ensembles of ETMC gauge field configurations with $N_f=2$ dynamical flavours of
quarks.
All of them correspond to a fixed physical situation of $L\approx1.3$ fm and a pion mass of ca. 330
MeV.
As we have argued above, our results for $\gamma_m$ can still be considered to be infinite-volume
ones (if $M\gtrsim500$ MeV) and pertaining to the chiral limit.
Our aim is now to relate them to continuum PT.
To achieve this, we have to take the continuum limit of lattice results.

\begin{figure}
\begin{center}
\includegraphics
[width=0.6\textwidth,angle=270]
{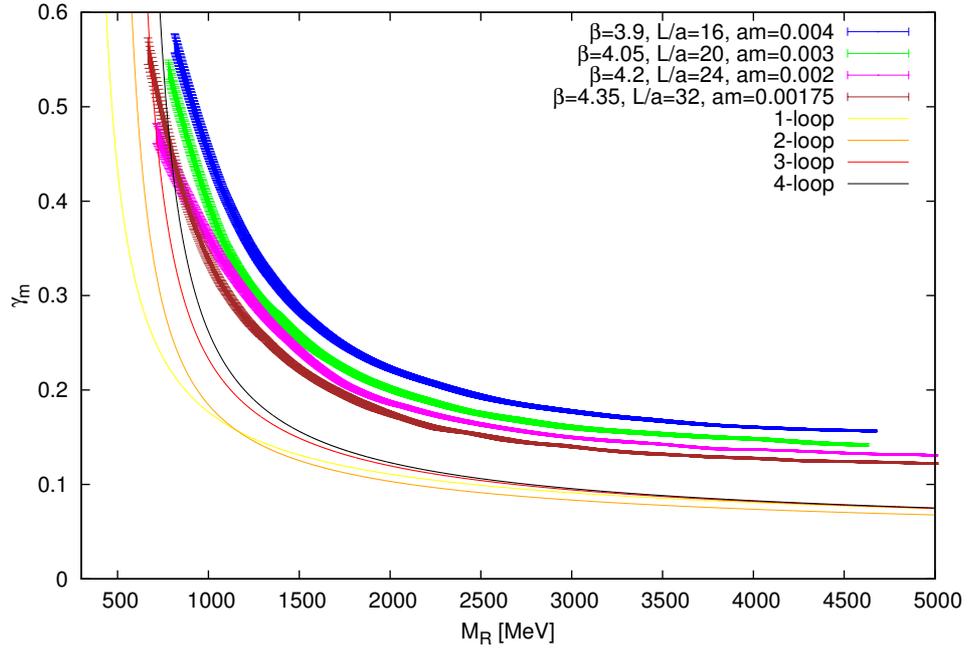}
\caption{Results for all 4 lattice spacings. Renormalized spectral threshold $M_R$.}
\label{fig:all_renorm}
\end{center}
\end{figure}

Fig.~\ref{fig:all_unrenorm} shows the values of $\gamma_m(M)$ vs. bare threshold parameter $M$ for all
4 lattice spacings.
We observe a clear dependence of the results on the lattice spacing.
In order to take the continuum limit, we have to renormalize the scale $M$ according to
Eq.~\eqref{eq:ren}.
In addition, to obtain the values of $\gamma_m(M_R)$ at arbitrary scales $M_R$ (and not only
the discrete set related to the values of $M$ chosen for the computation of the mode
number), we perform quadratic interpolation.
The outcome of renormalization of $M$ and interpolation is shown in Fig.~\ref{fig:all_renorm}.
The plot also shows the quark mass anomalous dimension $\gamma_m(a_s(\mu))$ 
(since we renormalize with $Z_P^{-1}(\mu\!=\!M_R)$, we can identify $\mu$ with $M_R$) in perturbation
theory, at different orders, with $\Lambda^{(2)}_{\MSb}=315$ MeV \cite{Jansen:2011vv} (this value is
always used in comparisons to PT, its quoted error -- 30 MeV -- is not reflected in the PT curves).
The difference between 3-loop and 4-loop results becomes small around 1.5-2 GeV, indicating that
these are the smallest values of $M_R$ where meaningful comparison of PT and lattice
is possible.

\begin{figure}
\begin{center}
\includegraphics
[width=0.6\textwidth,angle=270]
{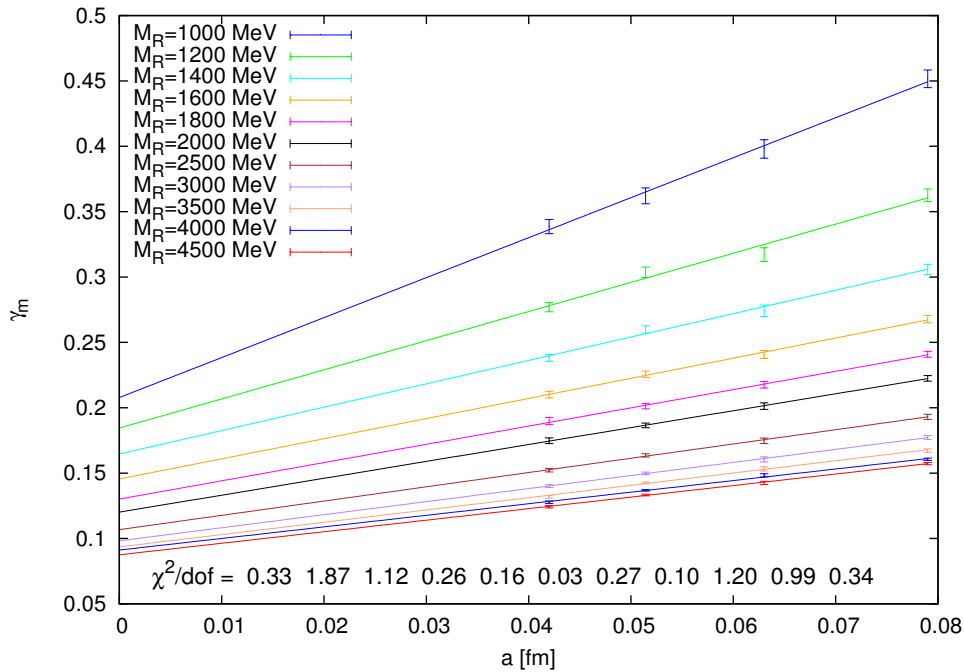}
\caption{Continuum limit extrapolations (Strategy 1) of $\gamma_m(M_R)$ at fixed $M_R$ (between 1
and 4.5 GeV). We
also give $\chi^2/{\rm d.o.f.}$ of the fits -- the values from left to right correspond to
increasing values of $M_R$, as indicated in the key.}
\label{fig:cont}
\end{center}
\end{figure}

\begin{table}[t!]
  \centering
\caption{\label{tab:cont}Continuum limit of the quark mass anomalous dimension $\gamma_m(M_R)$ at
different values
of $M_R$ and its error decomposition: statistical error, error originating from the value of lattice
spacing in physical units, error coming from $Z_P$. We also give 3- and 4-loop values of
$\gamma_m(a_s(M_R))$
\cite{Chetyrkin:1997dh,Vermaseren:1997fq}.}
  \begin{tabular}{clcc}
    $M_R$ & cont.limit. $\gamma_m(M_R)$  & \multicolumn{2}{c}{$\gamma_m(a_s(M_R))$}\\
    $[$MeV$]$ & (stat.)($\Delta a$)($\Delta Z_P$) & 3-loop & 4-loop\\
\hline
1000 & 0.208(13)(42)(18) & 0.2330 &  0.2619 \\
1500 & 0.155(6)(22)(10) & 0.1491 & 0.1563\\
2000 & 0.120(5)(14)(6)& 0.1197 &  0.1231 \\
2500 & 0.107(3)(11)(4)& 0.1041 &  0.1062\\
3000 & 0.098(3)(9)(3)& 0.0942 & 0.0956 \\
3500 & 0.094(3)(7)(3) & 0.0873 & 0.0883\\
4000 & 0.091(2)(6)(2)& 0.0820 & 0.0828\\
4500 & 0.087(2)(6)(2)& 0.0780 & 0.0786\\
  \end{tabular}
\end{table}

\begin{figure}
\begin{center}
\includegraphics
[width=0.6\textwidth,angle=270]
{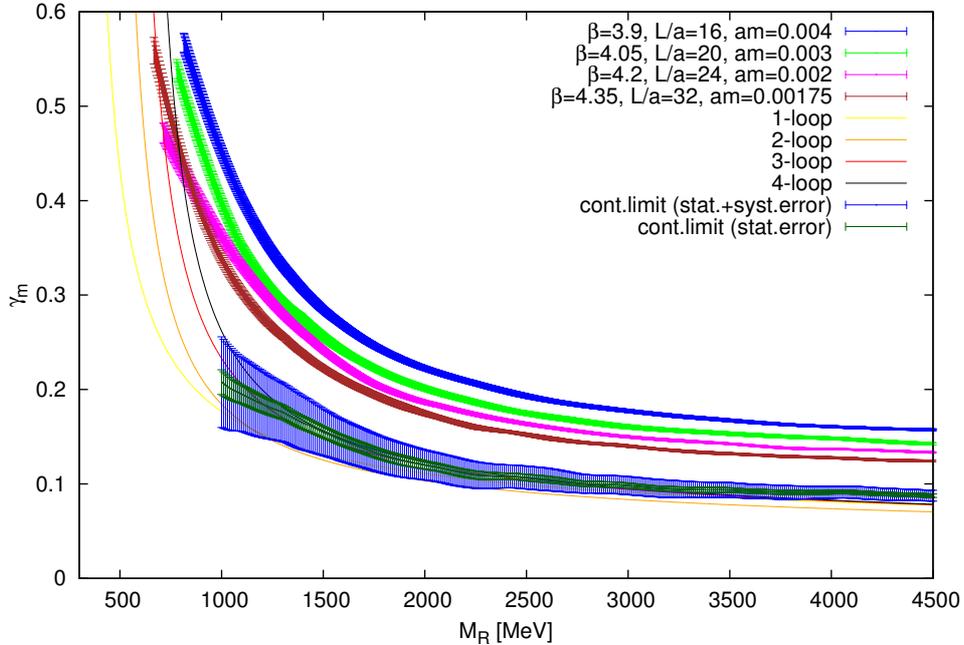}
\caption{Comparison of the quark mass anomalous dimension extracted from the lattice to
perturbation theory (for Strategy 1). For the continuum limit results, we show the statistical
error and the total one, i.e. the combined statistical error, the error originating from lattice
spacing value in physical units and the error coming from $Z_P$ (see Tab.~\ref{tab:setup} and
Sec.~\ref{sec:renorm} for more details).}
\label{fig:all_cont}
\end{center}
\end{figure}

The continuum limit extrapolations are shown in Fig.~\ref{fig:cont}. In all cases, the scaling is
consistent with $\mathcal{O}(a)$ cut-off effects.
The results in the continuum limit are also shown in Tab.~\ref{tab:cont}, for selected values
of $M_R$ and in Fig.~\ref{fig:all_cont} for the whole interval of $M_R$ between 1 and 4.5 GeV.
In Tab.~\ref{tab:cont}, we also give values of the quark mass anomalous dimension in perturbation
theory. 
We observe very good agreement between the lattice results (continuum-extrapolated) and perturbative
values in the range between 1 and around 3.5-4 GeV.
As we discussed above, the lower limit of $M$ that yields a reliable result from the point of view of
finite volume effects and the importance of the term $\nu_0(m)$ corresponds to $M\approx500$-600 MeV,
i.e. $M_R\approx1.3$-1.5 GeV.
In addition, the difference between 3- and 4-loop perturbative values below 1.5 GeV suggests that the
good agreement in the interval between 1 and 1.5 GeV should not be taken very seriously.
The upper limit of $M_R$ that can be used to simulate on the lattice is related to the lattice
cut-off. In our case, the inverse lattice spacings correspond to cut-offs of 2 to 4 GeV -- hence,
above these
values, one expects enhanced discretization effects.
This explains the observation that above ca. 4 GeV the lattice results yield $\gamma_m$ above the one
of PT.
The agreement can be regained by an inclusion of higher-order $\mathcal{O}(a^2)$ cut-off effects in
the fits, however at a price of a significantly increased error of the fits.

As a check of robustness of our results, we have performed three further checks, described in
appendices.

First, we checked the influence of including the term $\nu_0(m)$ in the fits on our continuum limit
extrapolations -- see Appendix \ref{sec:nu0}.
Here we only state the conclusion -- this influence is noticeable only for relatively small values of
$M_R$.
Depending on the value of $\nu_0(m)$, we observe agreement between lattice and PT
starting at $M_R$ between ca. 1100 and 1500 MeV for $\nu_0(m)$ in the range between 10 and
40.

Second, we followed the approach explained in Sec.~\ref{sec:fit} (fits of Eq.~\eqref{eq:nu7}) to
extract the ``artefact'' anomalous dimension $\Gamma_m(M_R)$ -- see Appendix \ref{sec:Gamma}.
We observed that the latter, as expected, is always compatible with zero in the continuum limit.

Third, to confirm that using the non-perturbative running of $Z_P$ yields compatible results with the
one using perturbative running of $Z_P$, we have repeated the procedure of this section applying the
latter, i.e. 4-loop evolution shown in Fig.~\ref{fig:ZP} -- see Appendix \ref{sec:ZPpert}.
As expected, the results are fully compatible, which confirms that the non-perturbatively evolved
$Z_P$ differs from the perturbatively evolved one only by cut-off effects (provided, of course, that
one stays at scales where PT is applicable).

\subsection{Continuum limit -- Strategy 2}
\label{sec:cont2}

\begin{figure}
\begin{center}
\includegraphics
[width=0.6\textwidth,angle=270]
{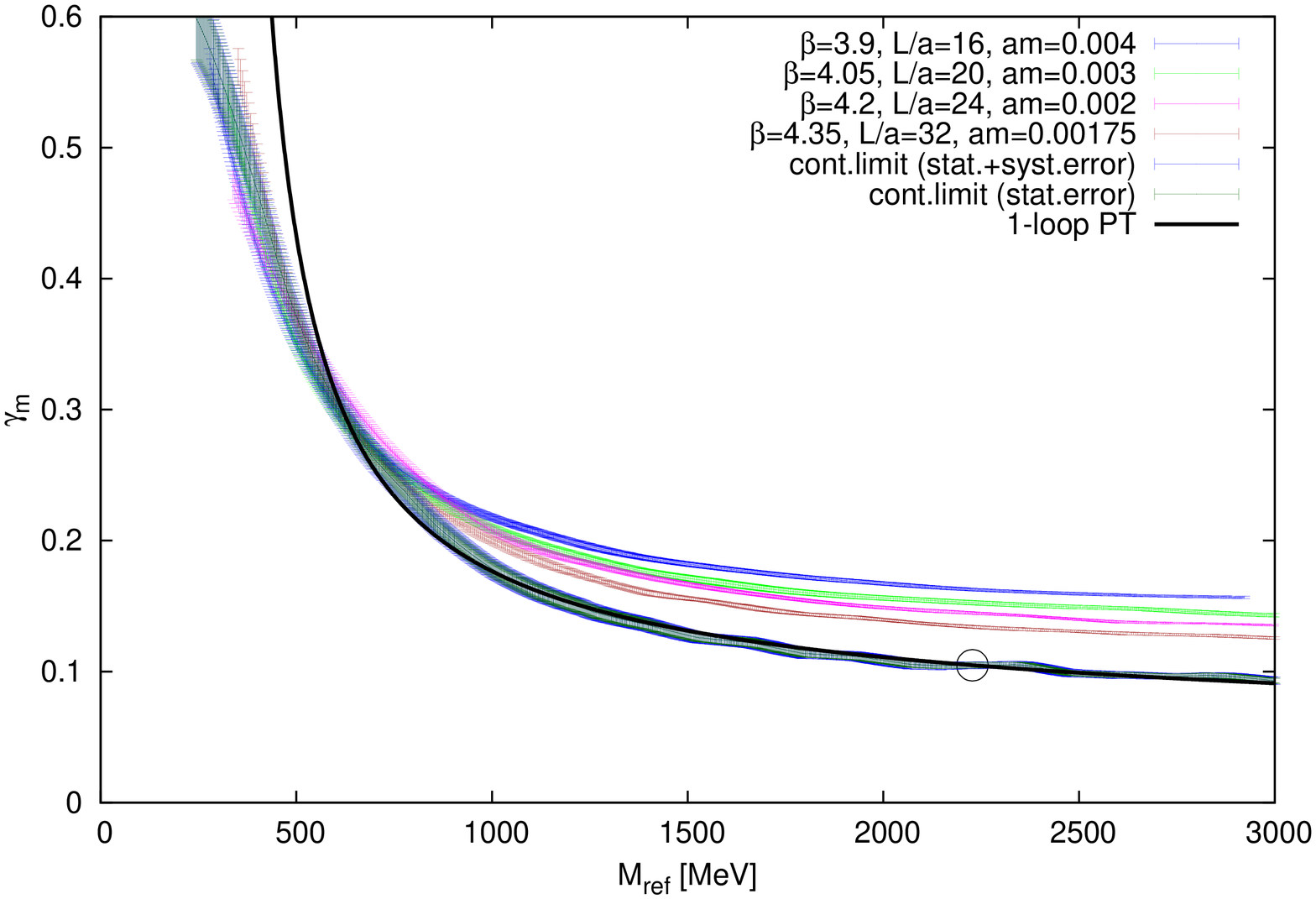}
\includegraphics
[width=0.6\textwidth,angle=270]
{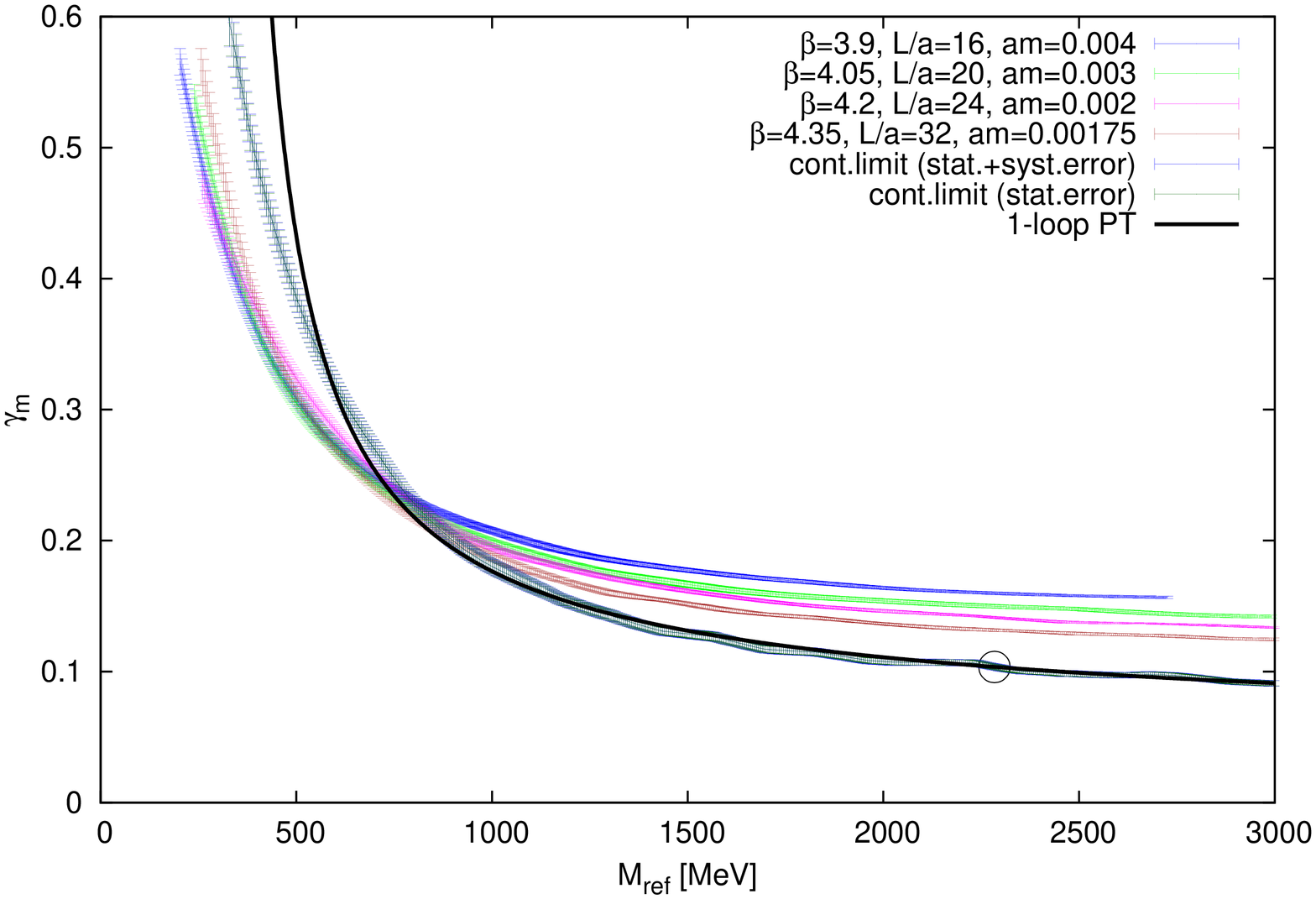}
\caption{Comparison of the quark mass anomalous dimension extracted from the lattice to
perturbation theory (for Strategy 2). The upper plot shows the case of $\beta_{\rm ref}=3.9$,
while the lower one of $\beta_{\rm ref}=4.35$. For the continuum limit results, we show the
statistical
error and the total one, i.e. the combined statistical error and the error originating from the
uncertainty in $r_0/a$ (see Tab.~\ref{tab:setup}).
The point of matching to PT is marked with a circle.}
\label{fig:all_cont2}
\end{center}
\end{figure}

We now present results of the other strategy of analysis, in which the values of $Z_P$ for different
ensembles are not needed.
Instead, one employs a rescaling procedure proposed in Ref.~\cite{Cheng:2013eu} and described in
Sec.~\ref{sec:str2}.
The choice of the reference value $\beta_{\rm ref}$ is arbitrary.
We will show results obtained for $\beta_{\rm ref}=3.9$ and $\beta_{\rm ref}=4.35$,
presented in Fig.~\ref{fig:all_cont2}.
In both cases we perform the matching to PT around 2.25 GeV, where 1-loop
PT yields $\gamma_m\approx0.105$ ($\Lambda^{(2)}_{\MSb}=315$ MeV \cite{Jansen:2011vv}).
In this way, the continuum extrapolated lattice result agrees with continuum PT at
2.25 GeV and thus the lattice eigenvalue $a_{\rm ref}M_{\rm ref}=0.8$ (for $\beta_{\rm ref}=3.9$) or
$a_{\rm ref}M_{\rm ref}=0.436$ (for $\beta_{\rm ref}=4.35$) corresponds to approx. 2.25 GeV in the
continuum.
Having performed this matching, the continuum values obtained from the lattice should agree with the
ones of PT, i.e. the scale dependence of the anomalous dimension should agree with
the one of PT for some range of scales.
We find this is really the case for the range of between around 1 GeV (or even somewhat below) and 3
GeV.
As discussed above, the upper bound is due to cut-off effects, while the lower one is set by
physical effects -- non-perturbative effects, in particular spontaneous chiral symmetry breaking
effects -- therefore the agreement below ca. 1.5 GeV can be coincidental.

In addition, strategy 2 gives independent estimates of the lattice spacings.
Working at $\beta_{\rm ref}=3.9$, we know that $a_{\beta=3.9}M=0.8$ corresponds to 2.25 GeV and
hence we obtain an estimate $a_{\beta=3.9}=0.071(7)(7)$ fm, where the first error is the combined
statistical and systematic error coming from the estimate of $\gamma_m$ in the continuum limit (i.e.
it combines the error of extraction of $\gamma_m(M)$ for all 4 ensembles, the errors of $r_0/a$ needed
to combine different lattice spacings and the error of the continuum extrapolation) and the second one
is an estimate of effects of using only 1-loop PT \footnote{Note that while a 4-loop
value of $\gamma_m$ is available in the $\MSb$ scheme, the lattice extracted value is not in this
scheme and hence only matching to 1-loop PT can be performed, since $\gamma_m$ is
universal at one loop. Nevertheless, some feeling for the size of higher order effects can be obtained
by comparing $\gamma_m$ at one and four loops in the $\MSb$ scheme, which amounts to a
difference of approx. 10 \% at the considered energy scale.}.
Using $\beta_{\rm ref}=4.35$, one obtains $a_{\beta=4.35}=0.038(1)(4)$ fm.
Both estimates of the lattice spacing agree within error with the ones quoted in
Tab.~\ref{tab:setup}, coming from chiral perturbation theory fits of the quark mass dependence of the
pseudoscalar meson mass and decay constant.
This agreement is reassuring, although the uncertainties are too large to be conclusive.

\subsection{Determination of $\Lambda^{(2)}_{\MSb}$}
The implicit scale dependence of the quark mass anomalous dimension is governed by the running of the
strong coupling constant, which, in turn, depends on the value of the $\Lambda$-parameter of the
underlying theory. The method analyzed in this paper allows to obtain the anomalous dimension with good
precision over a wide range of scales and hence, as will be demonstrated below, makes it possible to
determine this $\Lambda$-parameter. Specifically, it will be the $\Lambda$-parameter of 2-flavour QCD,
expressed in the $\MSb$ renormalization scheme -- hence we will denote it by $\Lambda^{(2)}_{\MSb}$.

\begin{figure}
\begin{center}
\includegraphics
[width=0.6\textwidth,angle=270]
{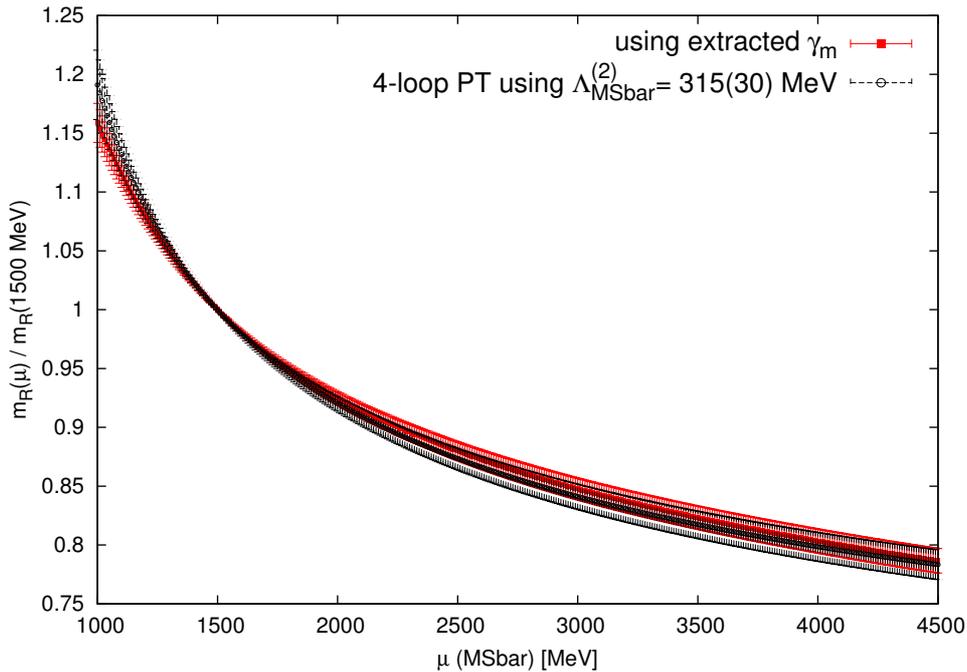}
\caption{The $\MSb$ scale dependence of the ratio of renormalized quark masses
$m_R(\mu)/m_R(\mu_{\rm ref})$, with $\mu_{\rm ref}=1.5$ GeV, determined non-perturbatively from the
extracted values of $\gamma_m$ and perturbatively \cite{Chetyrkin:1997dh,Vermaseren:1997fq} at 4-loops,
with $\Lambda^{(2)}_{\MSb}$=315(30) MeV \cite{Jansen:2011vv}.}
\label{fig:Lambda1}
\end{center}
\end{figure}

\begin{figure}
\begin{center}
\includegraphics
[width=0.6\textwidth,angle=270]
{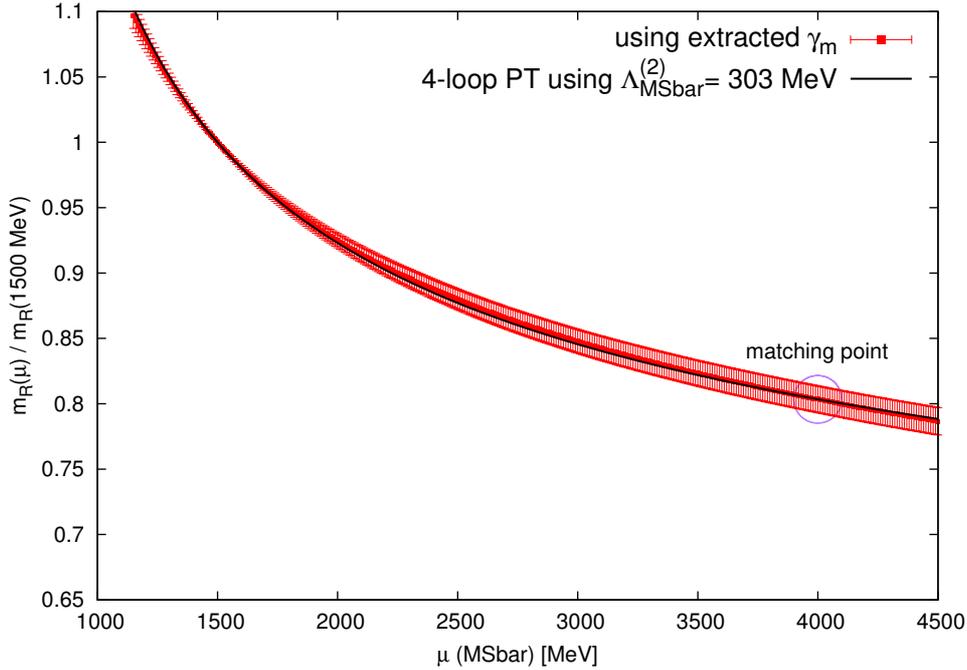}
\caption{The $\MSb$ scale dependence of the ratio of renormalized quark masses
$m_R(\mu)/m_R(\mu_{\rm ref})$,
with $\mu_{\rm ref}=1.5$ GeV, determined non-perturbatively from the extracted values of $\gamma_m$ and
perturbatively \cite{Chetyrkin:1997dh,Vermaseren:1997fq} at 4-loops, with
$\Lambda^{(2)}_{\MSb}$=303 MeV. The latter value is the result of matching the lattice result and
4-loop PT at $\mu_{\rm matching}=4$ GeV. See text for details.}
\label{fig:Lambda2}
\end{center}
\end{figure}

We start by showing the plot of the scale dependence of the ratio of renormalized quark masses
$m_R(\mu)/m_R(\mu_{\rm ref})$, with the chosen value of $\mu_{\rm ref}=1.5$ GeV (Fig.~\ref{fig:Lambda1}).
The two curves correspond to:
\begin{itemize}
\item the ratio $m_R(\mu)/m_R(\mu_{\rm ref})$ determined non-perturbatively from the
extracted (continuum extrapolated) values of $\gamma_m$ (using \eqref{eq:ZPrunning} with $\mu_2$ and
$\mu_1$ differing by 10 MeV),
\item 4-loop PT \cite{Chetyrkin:1997dh,Vermaseren:1997fq}, where the strong coupling
constant was computed using $\Lambda^{(2)}_{\MSb}$=315(30) MeV \cite{Jansen:2011vv} (value
corresponding to a lattice determination by ETMC, from the matching of $Q\bar{Q}$ static potential to
PT, here the uncertainty of $\Lambda^{(2)}_{\MSb}$ is reflected in the plot).
\end{itemize}
We observe very good agreement of the two curves within errors. The error of $m_R(\mu)/m_R(\mu_{\rm
ref})$ determined from $\gamma_m$ is comparable to the uncertainty of the PT curve
related to the uncertainty of the used value of $\Lambda^{(2)}_{\MSb}$.
Although the curves are compatible, it can be noticed that their shape is slightly
different, i.e. the latter is a bit steeper, suggesting that the value of $\Lambda^{(2)}_{\MSb}$
resulting from the former is actually smaller than 315 MeV (but compatible within errors).

To find the value of $\Lambda^{(2)}_{\MSb}$ that gives the best agreement between $\gamma_m$-extracted
$\Lambda^{(2)}_{\MSb}$ and the one from PT, we employ the following procedure.
Having chosen a reference scale $\mu_{\rm ref}$, we impose the matching condition at some
chosen scale $\mu_{\rm matching}$:
\begin{equation}
\frac{m_R(\mu_{\rm matching})}{m_R(\mu_{\rm ref})}\Big|_{\text{\normalsize from $\gamma_m$}}
=
\frac{m_R(\mu_{\rm matching})}{m_R(\mu_{\rm ref})}\Big|_{\text{\normalsize PT with
$\Lambda^{(2)}_{\MSb}=\overline{\Lambda}$}}, 
\end{equation} 
which defines the value $\overline{\Lambda}$, such that the ratio $m_R(\mu)/m_R(\mu_{\rm ref})$ extracted
from $\gamma_m$ and computed in 4-loop PT with $\Lambda^{(2)}_{\MSb}=\overline{\Lambda}$
agrees for $\mu=\mu_{\rm ref}$ (by construction) and for $\mu=\mu_{\rm matching}$ (by matching).
Of course, if the method to extract $\gamma_m$ from the Dirac operator spectrum is valid, the agreement
should also ensue for all other scales (down to $\mu$ such that 4-loop PT is no
longer acceptable).

The above procedure is illustrated in Fig.~\ref{fig:Lambda2}. As before, $\mu_{\rm ref}=1.5$ GeV, while
$\mu_{\rm matching}$ is chosen to be 4 GeV. The value of $\Lambda^{(2)}_{\MSb}$ that leads to the
fulfillment of the matching condition is 303 MeV. It is also striking that the values of
$m_R(\mu)/m_R(\mu_{\rm ref})$ from $\gamma_m$ and from PT agree remarkably well for the
whole range of scales $\mu$ (for which the continuum-extrapolated lattice data for $\gamma_m$ are
available).

It now remains to establish the robustness of the result. We have performed a thorough error analysis.
The considered sources of uncertainty are: statistical error, uncertainty from the input value of
$Z_P^{\MSb,\mu_2}$ (from Tab.~\ref{tab:zp}, i.e. no perturbative running of the non-perturbatively found
RI-MOM/X-space values is performed), error from relative and absolute scale setting, uncertainty from the
arbitrary choice of $\mu_{\rm ref}$ and $\mu_{\rm matching}$ and, finally, uncertainty from neglecting
higher order terms in PT.

\begin{table}[t!]
\begin{center}
\caption{Error budget for our computation of $\Lambda^{(2)}_{\MSb}$. The total systematic error comes
from combining the individual ones in quadrature. All values in MeV. See text for more details.}
\begin{tabular}{lc}
\hline
\hline
central value [MeV] & 303\\
\hline
$Z_P^{\MSb,\mu_2}$ & 16\\
relative scale setting & 12\\
absolute scale setting & 6\\
choice of $\mu_{\rm ref}$ & 15\\
choice of $\mu_{\rm matching}$ & 4\\
higher orders of perturbation theory & 1\\
\hline
total systematic & 25\\
statistical & 13 \\
\hline
\end{tabular}
\label{tab:errors}
\end{center}
\end{table} 

To estimate the statistical error, a bootstrap with blocking (to account for possible
autocorrelations) procedure was performed. The bootstrap procedure was also carried out to find the
propagation of the error from the input values of $Z_P^{\MSb,\mu_2}$ and from scale setting. For the
latter, effects of relative scale setting and absolute scale setting were separated. The relative scale
setting uncertainty was estimated by generating artificial bootstrap samples of the continuum limit of
$\gamma_m$ performed at different values of $r_0/a$, Gaussian distributed around the central values
given in Tab.~\ref{tab:setup}, with errors as given in parentheses. For the effects of absolute scale
setting, the computation of $m_R(\mu)/m_R(\mu_{\rm ref})$ was repeated for all values of the lattice
spacings in physical units (Tab.~\ref{tab:setup}) changed by 10\%. To estimate the effects of an
arbitrary choice of $\mu_{\rm ref}$ and $\mu_{\rm matching}$, we again repeated the whole calculation
procedure for different values of these two scales. Finally, for the estimate of neglecting the higher
order
terms of PT, we compared the results from 3-loop and 4-loop PT.

The results of the error analysis are gathered in Tab.~\ref{tab:errors}. The relative statistical error
amounts to around 4\%, while the total systematic error is around 8\%. The dominating sources of the
latter are: the uncertainty of the values of $Z_P$ (i.e. mostly systematic errors of the non-perturbative
extraction of $Z_P$ in the RI-MOM or X-space schemes), which are an external input to the present
analysis and the choice of the reference scale $\mu_{\rm ref}$ and to a lesser extent the uncertainty of
the input values of $r_0/a$. Note that the absolute scale setting and the choice of $\mu_{\rm matching}$
lead to very small systematic uncertainties of $\Lambda^{(2)}_{\MSb}$, while the effects from neglecting
higher orders of PT are essentially negligible.

Finally, we quote:
\begin{equation}
 \Lambda^{(2)}_{\MSb}=303(13)(25)\;\text{MeV},
\end{equation} 
where the first error is statistical and the second one systematic. This can be compared to other
recent determinations in the literature, e.g. 315(30) MeV (ETMC, Ref.~\cite{Jansen:2011vv}), 331(21) MeV
(ETMC, Ref.~\cite{Karbstein:2014bsa}) and
310(20) MeV (ALPHA Collaboration, Ref.~\cite{Fritzsch:2012wq}). We find very good agreement with these
values and a similar total error. For a more thorough comparison with earlier determinations, see the
conclusions of Ref.~\cite{Jansen:2011vv}.
Note, however, that there is a tension between the two ETMC values and the ALPHA value, when the result
is expressed as a dimensionless product $r_0\Lambda^{(2)}_{\MSb}$ (0.658(55) and 0.692(31) vs. 0.789(52),
respectively).
The analysis presented here is independent of the value of $r_0$ in physical units \footnote{The
continuum limits of $\gamma_m$ at different scales $M_R$ depend on the relative lattice spacings, given
by the dimensionless quantities $r_0/a$ and then matching to PT depends on absolute
scale setting, but the results for $\Lambda^{(2)}_{\MSb}$ in MeV differ by only around 2\% if the
absolute scale is varied over 10\%.} and hence we do not give a value of $r_0\Lambda^{(2)}_{\MSb}$.

Another comparison with the results from the ALPHA Collaboration concerns the curve
$m_R(\mu)/m_R(\mu_{\rm ref})$ (using extracted $\gamma_m$), shown in Fig.~\ref{fig:Lambda1}. This curve
gives the same information as
the curve $\overline{m}(\mu)/M$ in Fig.~4 of Ref.~\cite{DellaMorte:2005kg}, where $\overline{m}$ is
the renormalized quark mass and $M$ is the RG-invariant quark mass. Although the curves are expressed in
different renormalization schemes ($\MSb$ and Schr\"odinger functional, respectively), their
respective agreement with PT suggests also the mutual agreement between the results of this paper and
Ref.~\cite{DellaMorte:2005kg}. 

\section{Conclusions}
\label{Sec:conclusions}
The main conclusion of this work can be formulated in the following way -- there exists a window in
which the quark mass anomalous dimension extracted from the lattice and extrapolated to the continuum
limit agrees well with continuum perturbation theory.
Although the relation between the quark mass anomalous dimension and the scaling of the mode number can
be shown in perturbation theory \cite{Cheng:2013bca}, it was a priori unclear whether the above mentioned
window exists with presently simulated lattices.

One of the main aims of this work was to investigate whether the scaling of the
mode number for intermediate eigenvalues of the Hermitian Dirac operator can be described in QCD in
a similar way as in conformal field theories with an infrared fixed point.
We found that the answer is indeed positive -- provided one stays well
above the non-perturbative regime.
We have employed two analysis strategies and overall the results obtained using both of them are
consistent.
In both cases, we observe very good agreement with continuum perturbation theory in some range of
energy scales.
What is worth emphasizing is that both strategies correctly predict the scale dependence of the quark
mass anomalous dimension, although they obtain it with different means -- either by converting the
lattice extracted value to the $\MSb$ scheme using $Z_P$ in this scheme, or by matching to
perturbation theory.

Taking into account all the systematic effects, we estimate that the above mentioned contact window
between the lattice and perturbation theory extends from around 1.5-2 GeV to 3.5-4 GeV at presently
used lattice spacings.
The lower limit -- ca. 1.5 GeV -- originates from two physical reasons.
\begin{itemize}
\item In the low-energy regime, QCD with $N_f=2$ flavours of quarks exhibits spontaneous chiral
symmetry breaking, signaled by a non-zero value of the chiral condensate.
In this regime, the spectral density does not tend to zero near the origin -- hence, its scaling
does not follow Eq.~\eqref{eq:spectral} and the mode number obtains an additive correction from the
non-zero value of spectral density at the origin.
Taking realistic values of this contribution, we estimated that the influence of spontaneous chiral
symmetry breaking regime extends to scales of $M_R\approx1.5$ GeV.
\item The lower end of the lattice window is also related to the applicability of perturbation theory
at low energies. Taking the difference of 3- and 4-loop perturbative results as a proxy of this
applicability, it can be estimated that 4-loop perturbation theory formula for the quark mass
anomalous dimension can be trusted above around 1.5-2 GeV (with strong coupling constant
$\alpha_s\approx0.3$ at this scale).
\end{itemize}
The upper limit of the window is related to the values of lattice spacings in present-day simulations
and can be, in principle, improved by using finer lattice spacings.
On the other hand, the lower limit is dictated by physics, i.e. non-perturbative effects, in
particular originating from spontaneous breaking of chiral symmetry, that dominate at energies below
the lower limit.

The computation of the scale dependence of the quark mass anomalous dimension allowed also for an
extraction of the $\Lambda$-parameter of 2-flavour QCD. The obtained value:
\begin{equation}
 \Lambda^{(2)}_{\MSb}=303(13)(25)\;\text{MeV},
\end{equation} 
where the first error is statistical and the second one systematic, agrees well with earlier
determinations, in particular the recent ones by ALPHA \cite{Fritzsch:2012wq} and ETM
\cite{Jansen:2011vv,Karbstein:2014bsa} collaborations.

\begin{acknowledgments}
I thank the European Twisted Mass Collaboration for the generation
of gauge field configurations used in this project.
The effective realization of this project was possible because of the earlier work regarding the
chiral condensate from the mode number, in particular the implementation of the mode number
computation in the tmLQCD code, done in collaboration with Elena Garcia Ramos and Karl Jansen.
I am especially grateful to Gregorio Herdoiza for several illuminating discussions, suggestions and
comments.
I acknowledge very useful correspondence and discussions with Anqi Cheng, Anna Hasenfratz and David
Schaich.
I also thank Konstantin G. Chetyrkin, Vincent Drach, Karl Jansen, Piotr
Korcyl, Johann H. K\"uhn, Elisabetta Pallante, Giancarlo Rossi and Stefan Sint for discussions and/or
suggestions.
This work has been supported in part by Foundation for Polish Science fellowship ``Kolumb''
and by the DFG Sonderforschungsbereich/Transregio SFB/TR9. 
The computations for this project were performed on SuperMUC at LRZ Munich, the PC cluster in Zeuthen
and Poznan Supercomputing and Networking Center (PCSS). 
I thank these computer centers and their staff for all technical advice and help.
\end{acknowledgments}

\begin{figure}
\begin{center}
\includegraphics
[width=0.6\textwidth,angle=270]
{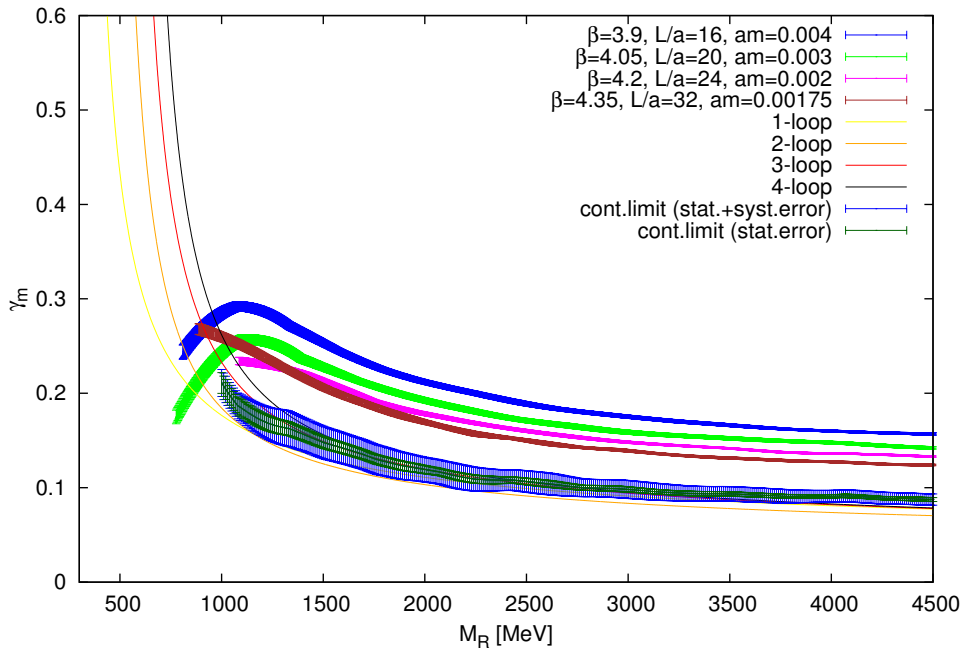}
\caption{Comparison of the quark mass anomalous dimension extracted from the lattice to
perturbation theory. Results of alternative analysis, with the term $\nu_0(m)$ (set to 20) included in
the fits of the mode number.}
\label{fig:all_cont_nu0}
\end{center}
\end{figure}

\appendix

\section{Continuum limit extrapolation with the term $\nu_0(m)$ included in the fits}
\label{sec:nu0}
The influence of the term $\nu_0(m)$ on extracted values of the anomalous dimension was discussed in
Sec.~\ref{sec:procedure}.
In Fig.~\ref{fig:all_cont_nu0}, we show the outcome of an alternative continuum limit analysis
-- assuming that $\nu_0(m)=20$ is included in Eq.~\eqref{eq:nu3}. This plot should be compared to
Fig.~\ref{fig:all_cont}, which shows the results in the case $\nu_0(m)=0$.
Below $M_R\approx1$ GeV, the effect is huge (with a non-physical maximum of $\gamma_m$ around this
scale), signaling a total breakdown of the scaling formula for the mode number.
The difference in the obtained values of $\gamma_m(M_R)$ at a non-vanishing lattice spacing reaches to
a renormalized threshold parameter $M_R$ range of 1.5-2 GeV.
However, the difference in continuum limit extrapolated anomalous dimension can not be seen above
$M_R=1$ GeV (note, however, that the continuum limit extrapolation breaks down below about 1.2 GeV). 

This analysis can also be repeated for different values of $\nu_0(m)$. Finally, we can compare
results with $\nu_0(m)=0$, 10, 20, 30 and 40. Our conclusions are qualitatively the same in all cases
-- extrapolating to the continuum limit, the results are always compatible above $M_R\approx1.5$ GeV,
even using the conservative estimate $\nu_0(m)=40$.
Therefore, we conclude that 1.5 GeV is the lower limit for the applicability of the scaling
formula for the mode number.

\begin{figure}
\begin{center}
\includegraphics
[width=0.6\textwidth,angle=270]
{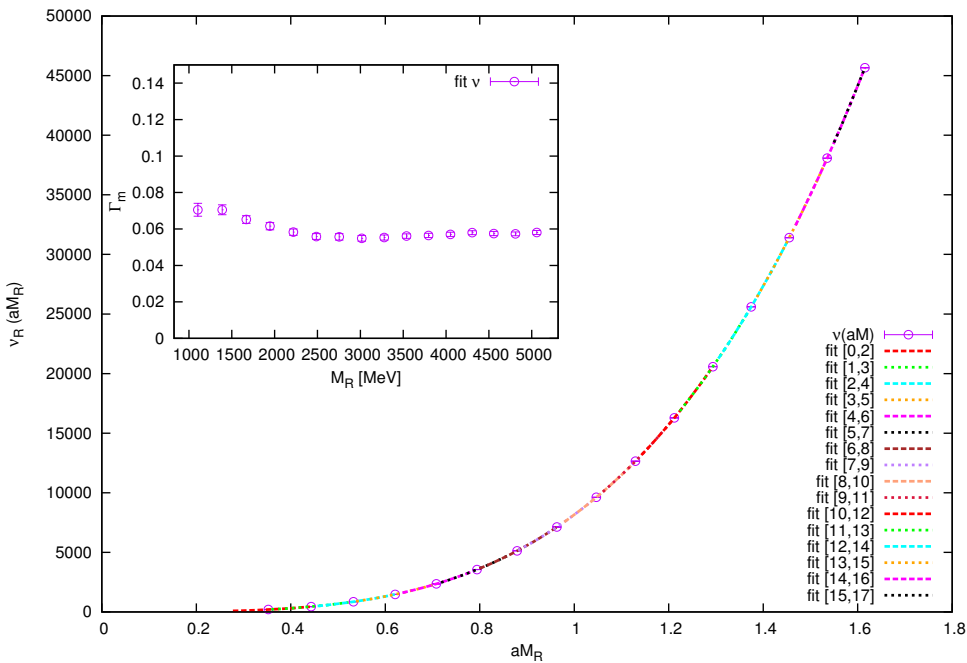}
\caption{Alternative approach to the scaling of the mode number -- fits of Eq.~\eqref{eq:nu7}.
Ensemble C30.20: $\beta=4.05$, $L/a=20$, $am=0.003$. The extracted $\Gamma_m$ should be
zero up to lattice artefacts.}
\label{fig:another}
\end{center}
\end{figure}

\section{``Artefact'' anomalous dimension $\Gamma_m$}
\label{sec:Gamma}

In Sec.~\ref{sec:fit}, we discussed an alternative approach for the analysis of the scaling of the
mode number, which consists in rewriting equation for the dependence of $\nu_R(M_R)$ on
$M_R$ at the renormalization scale $\mu=M_R\,$. The obtained Eq.~\eqref{eq:nu6} implies that
$\nu_R(M_R)\propto M_R^4$, i.e. the scaling exponent is independent of the quark mass
anomalous dimension $\gamma_m(M_R)$ and equals 4, up to cut-off effects.

\begin{figure}
\begin{center}
\includegraphics
[width=0.6\textwidth,angle=270]
{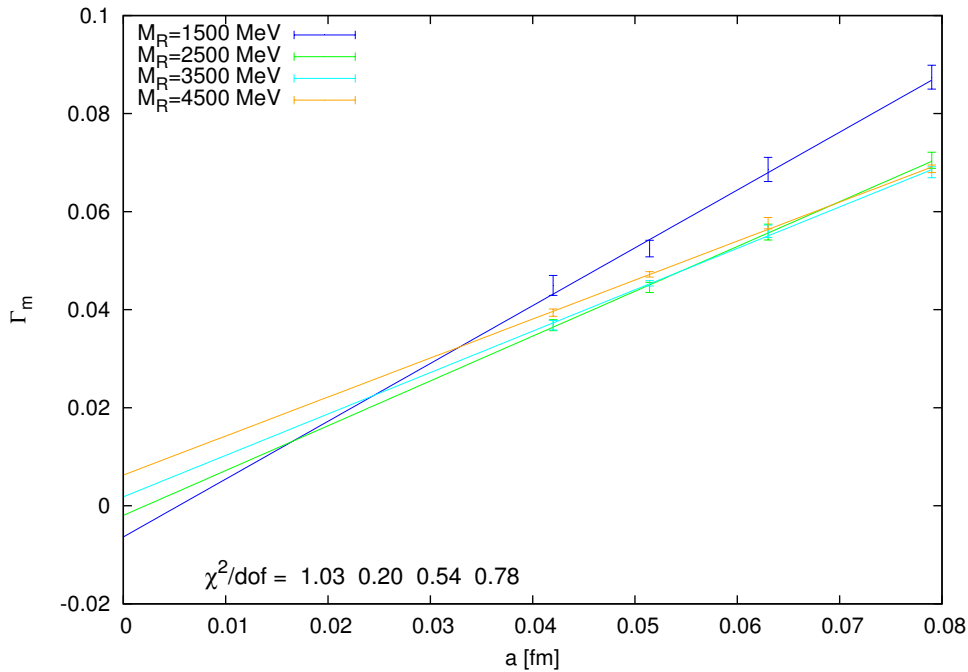}
\caption{Continuum limit extrapolations of the ``artefact'' anomalous dimension $\Gamma_m(M_R)$ at
fixed $M_R$ (between 1.5 and 4.5 GeV). We also give $\chi^2/{\rm d.o.f.}$ of the fits -- the values
from left to right correspond to increasing values of $M_R$, as indicated in the key.}
\label{fig:cont2}
\end{center}
\end{figure}

Here, we show the outcome of applying this alternative approach.
We fitted Eq.~\eqref{eq:nu7} to the data for $\nu_R(M_R)$ vs. $M_R$ and extracted the ``artefact''
anomalous dimension $\Gamma_m(M_R)$. An example for the ensemble C30.20 is given in
Fig.~\ref{fig:another}.
As expected, the ``artefact'' anomalous dimension is approximately constant in the whole range and
close to zero.
However, only a continuum limit analysis can show whether it is indeed zero.
The continuum extrapolations for selected values of $M_R$ (1.5 to 4.5 GeV) are shown in
Fig.~\ref{fig:cont2}.
In all cases, the obtained values in the continuum are compatible with zero and all fits have
$\chi^2/{\rm d.o.f.}\lesssim1$.
This fully confirms our expectations that $\nu_R(M_R)\propto M_R^{4+\mathcal{O}(a)}$ and gives
further indication of the self-consistency of the presented approach.

\section{Continuum limit extrapolation using perturbatively evolved $Z_P$}
\label{sec:ZPpert}

\begin{figure}
\begin{center}
\includegraphics
[width=0.6\textwidth,angle=270]
{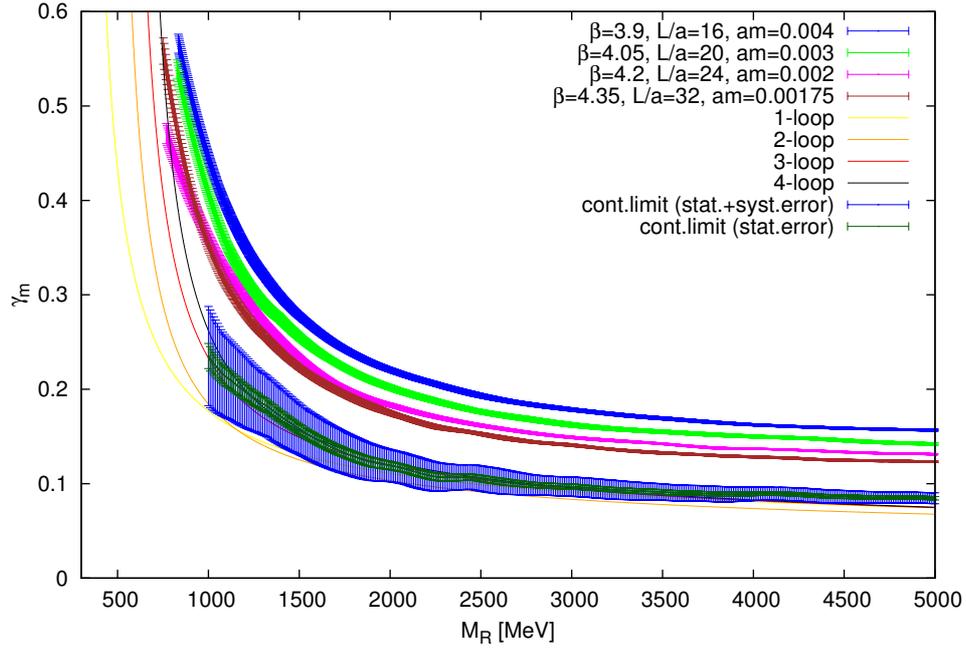}
\caption{Comparison of the quark mass anomalous dimension extracted from the lattice to
perturbation theory -- Strategy 1, using perturbative running of $Z_P$.
This plot should be compared with Fig.~\ref{fig:all_cont}, where non-perturbative running of $Z_P$ is
applied.
For the continuum limit
results, we show the statistical
error and the total one, i.e. the combined statistical error, the error originating from lattice
spacing value in physical units and the error coming from $Z_P$.}
\label{fig:ZPpert}
\end{center}
\end{figure}

As we have explained above, in order to compare the continuum limit of lattice extracted quark mass
anomalous dimension with PT prediction, it is essential that the threshold parameter
$M$ is renormalized with $Z_P$ that is evolved without using perturbative
expressions for $\gamma_m$.
However, as a check of self-consistency of the approach, we have performed the renormalization of $M$
using also perturbatively evolved $Z_P$.
The results of this check are shown in Fig.~\ref{fig:ZPpert}.
Minor differences between the cases of non-perturbatively and perturbatively evolved $Z_P$ are
observed only around 1 GeV, although they are still statistically insignificant.
We emphasize that this confirms that the curves showing perturbative and non-perturbative running of
$Z_P$ differ only by cut-off effects (in the regime where PT is applicable).

\begin{figure}
\begin{center}
\includegraphics
[width=0.34\textwidth,angle=270]
{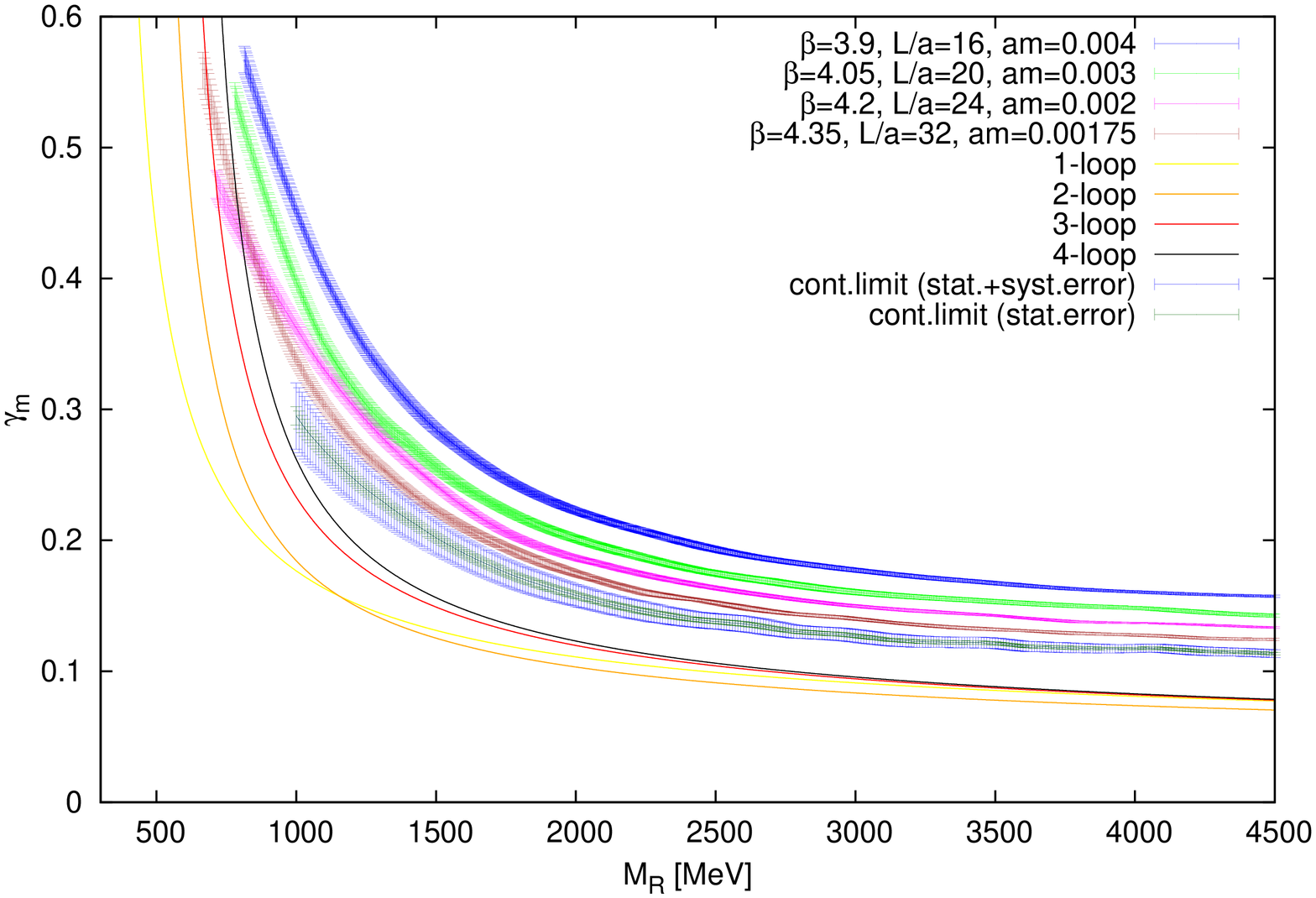}
\includegraphics
[width=0.34\textwidth,angle=270]
{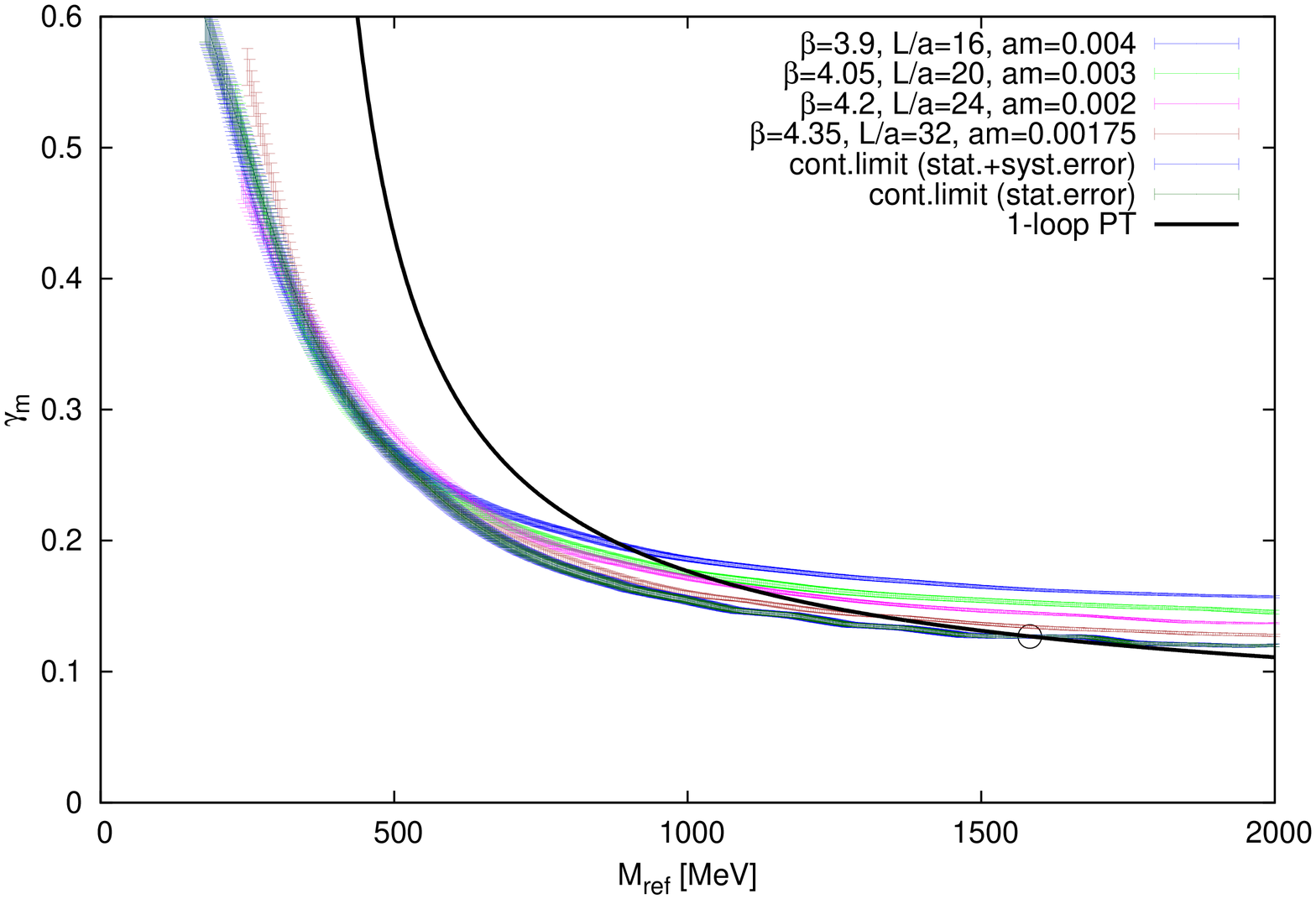}
\caption{Comparison of the quark mass anomalous dimension extracted from the lattice to
perturbation theory, assuming $\mathcal{O}(a^2)$ scaling towards the continuum. The left plot
shows
the results from Strategy 1 and the right plot from Strategy 2. In the latter, the matching point
is
marked with a circle.}
\label{fig:a2}
\end{center}
\end{figure}

\section{$\mathcal{O}(a)$ vs. $\mathcal{O}(a^2)$ scaling}
\label{sec:a2}

$\mathcal{R}_5$-parity even quantities computed
with twisted mass fermions at maximal twist are $\mathcal{O}(a)$-improved \cite{Frezzotti:2003ni}.
The situation is more complex with off-shell quantities in which contact terms can spoil the automatic
$\mathcal{O}(a)$-improvement -- an example of such quantity is the mode number.
However, it was shown in Ref.~\cite{Cichy:2013egr} that this does not happen for the case of the mode
number. 
Still, as we discussed in Sec.~\ref{sec:str1}, this does not imply that the quark mass anomalous
dimension extracted with the analyzed method is also $\mathcal{O}(a)$-improved.
In previous sections, we have performed the continuum limit extrapolations under the assumption that
$\mathcal{O}(a)$ effects can be present and indeed we found numerically that the coefficient of the
$\mathcal{O}(a)$-terms (at several values of $M_R$) is not consistent with zero.

Nevertheless, we performed some numerical checks of the continuum limit scaling of $\gamma_m$
\emph{assuming} $\mathcal{O}(a)$ effects are absent.
Since $\mathcal{O}(a^2)$ scaling involves much shorter extrapolations to the continuum, the
values in the continuum are very different from the ones assuming $\mathcal{O}(a)$ scaling.
The analogue of Fig.~\ref{fig:cont} assuming $\mathcal{O}(a^2)$ scaling shows much
inferior fits, i.e. the values of $\chi^2/{\rm d.o.f.}$ are much above 1.
However, taking errors into account, this test is not fully conclusive, as typically the values of
$\chi^2/{\rm d.o.f.}$ are only a factor of 2-3 worse than the ones obtained assuming
$\mathcal{O}(a)$ scaling and in some cases (for some values of $M_R$ or $M_{\rm ref}$) they are
even comparable.

The results of performing continuum extrapolations using Strategy 1 and Strategy 2 and assuming
$\mathcal{O}(a^2)$ scaling are shown in Fig.~\ref{fig:a2}.
Needless to say, both strategies lead to results in contradiction with PT in terms
of values predicted at a given physical scale (Strategy 1) or the scale dependence of the anomalous
dimension (Strategy 2; matching performed at around 1.6 GeV).
This further confirms that indeed $\mathcal{O}(a)$ leading cut-off effects are present in our data.

\bibliographystyle{jhep}
\bibliography{anom}	

\end{document}